\pdfoutput=1
\documentclass[article, 11 pt, notitlepage, nofootinbib, superscriptaddress, preprintnumbers, longbibliography, floatfix]{revtex4-2}
\usepackage{amsmath, amsfonts, amssymb, hyperref, xcolor, graphicx, xspace, enumitem,slashed,braket,cancel}
\usepackage[T1]{fontenc}
\definecolor{nicered}{rgb}{.7,.1,.1}
\definecolor{nicegreen}{rgb}{.1,.5,.1}
\definecolor{darkblue}{rgb}{0,0,.5}
\hypersetup{colorlinks, citecolor=nicegreen, linkcolor=nicered, urlcolor=darkblue}

\usepackage{setspace}
\usepackage{tikz}
\usetikzlibrary{arrows,shapes}
\usetikzlibrary{trees}
\usetikzlibrary{matrix,arrows} 				
\usetikzlibrary{calc}                       
\usetikzlibrary{positioning}				
\usetikzlibrary{calc,through}				
\usetikzlibrary{decorations.pathreplacing}  
\usepackage[tikz]{bclogo} 					
\usepackage{pgffor}						    
\usepackage{nicematrix}
\usetikzlibrary{decorations.pathmorphing}   
\usetikzlibrary{decorations.markings}
\usetikzlibrary{intersections}				

\newcommand{\delir}{{\delta_{\rm IR}}}
\DeclareMathOperator{\tr}{tr}
\begin{document}


\title{Functional Determinants for False Vacuum Decay}

\author{Pietro Baratella
\href{https://orcid.org/0000-0002-7891-8933}{
\includegraphics[scale=0.3]{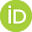}}}
\email{pietro.baratella@ijs.si}
\affiliation{Jo\v{z}ef Stefan Institute, Jamova 39, 1000 Ljubljana, Slovenia}

\author{Miha Nemev\v{s}ek
\href{https://orcid.org/0000-0003-1110-342X}{
\includegraphics[scale=0.3]{orcid_32x32.png}}}
\email{miha.nemevsek@ijs.si}
\affiliation{Jo\v{z}ef Stefan Institute, Jamova 39, 1000 Ljubljana, Slovenia}
\affiliation{Faculty of Mathematics and Physics, University of Ljubljana, Jadranska 19, 
1000 Ljubljana, Slovenia}

\author{Yutaro Shoji
\href{https://orcid.org/0000-0003-3932-9199}{
\includegraphics[scale=0.3]{orcid_32x32.png}}}
\email{yutaro.shoji@ijs.si}
\affiliation{Jo\v{z}ef Stefan Institute, Jamova 39, 1000 Ljubljana, Slovenia}

\author{\\Katarina Trailović
\href{https://orcid.org/0009-0000-2283-7408}{
\includegraphics[scale=0.3]{orcid_32x32.png}}}
\email{katarina.trailovic@ijs.si}
\affiliation{Jo\v{z}ef Stefan Institute, Jamova 39, 1000 Ljubljana, Slovenia}
\affiliation{Faculty of Mathematics and Physics, University of Ljubljana, Jadranska 19, 
1000 Ljubljana, Slovenia}

\author{Lorenzo Ubaldi
\href{https://orcid.org/0000-0002-9567-9719}{
\includegraphics[scale=0.3]{orcid_32x32.png}}}
\email{lorenzo.ubaldi@ijs.si}
\affiliation{Jo\v{z}ef Stefan Institute, Jamova 39, 1000 Ljubljana, Slovenia}

\date{\today}

\begin{abstract}
We derive simple expressions to regularise functional determinants from 
fluctuations of fields with spin 0, 1/2, and 1.
These are important for the precise dimensionful determination of false 
vacuum decay rates.
We work in $D = 4$ Euclidean dimensions and use familiar Feynman
diagrammatic techniques with a double expansion in interactions and masses,
together with dimensional regularisation in momentum space.
We Fourier transform to coordinate space and end up with a simple
regularisation prescription in terms of single integrals over the Euclidean radius of 
field-dependent masses and their derivatives.
Our results apply to models with an arbitrary scalar potential and with any number 
of scalars, fermions, gauge bosons and associated ghosts.
We exemplify this approach on the Standard Model with a streamlined calculation of 
the renormalisation and isolation of divergences in fluctuation determinants.
\end{abstract}

\maketitle

\newpage

\tableofcontents

%
%
\section{Introduction}

The quantitative study of bubble nucleation has applications in many areas of physics.
They range from the study of metastable states in thermal physics, to the computation of
the Standard Model (SM) electroweak vacuum lifetime, statistical and condensed 
matter physics.
If the system of interest is described in terms of a weakly coupled Quantum Field Theory 
(QFT), the standard procedure~\cite{Coleman:1977py, Callan:1977pt} for computing the 
rate of bubble formation applies.
It consists of first computing the bounce $\overline \phi$, which is a semi-classical 
scalar field configuration that extremises the action $S[\phi]$ 
and connects the false vacuum (FV) $\hat \phi$ with an escape point. 
Then one has to perform the path integral in a Gaussian approximation, expanding $S$ around 
the bounce and truncating to quadratic order in the fluctuations. 
The path integral sums up all the contributions from any field that couples to the bounce: 
scalars $\phi$, fermions $\psi$ and vectors (gauge bosons) $A_\mu$. 
This is a one-loop effect~\cite{Callan:1977pt} that boils down to computing the functional
determinant of the fluctuation operator $S''[\phi,\psi,A_\mu]$, which 
is the second variational derivative of the action. 
The inclusion of one-loop effects and renormalisation is needed for obtaining a finite and 
dimensionful nucleation rate, which is usually schematically written in the form
$\gamma = Ae^{-S[\bar\phi]}$, where the prefactor $A$ encodes the functional determinant.

A vast part of the literature on vacuum decay concentrates on the exponent and 
estimates the magnitude of $A$ based on dimensional analysis. 
This is not only because the priority is to get the exponent right, but
also because the functional determinant is a lot harder to compute~\cite{Callan:1977pt}.
Nonetheless, there has been some pioneering work on the latter, done in
two~\cite{Voloshin:1985id, Konoplich:1980au, Konoplich:1983mn, Kiselev:1975eq, 
Kiselev:1985er, Alford:1993ph, Alford:1993zf}, three~\cite{Strumia:1998nf, 
Strumia:1999fq,Munster:1999hr, Munster:2000qt, Munster:2003an, Moss:2001my}, and 
four dimensions~\cite{Konoplich:1985rir, Konoplich:1987yd, Garbrecht:2015oea}. 
Apart from Ref.~\cite{Moss:2001my}, in which the functional determinant is calculated
numerically for fermion fluctuations around a bounce solution approximated as a step 
function, the rest of the studies focused on scalar fluctuations. 
Most of them worked in the thin-wall approximation, in which the metastable and true 
vacua are almost degenerate.
Still for the scalar case, Refs.~\cite{Baacke:2003uw, Dunne:2005rt, Dunne:2006ct}
refined the study of functional determinants and provided prescriptions which work when 
departing from the thin-wall limit as well. 
In particular, Refs.~\cite{Dunne:2005rt, Dunne:2006ct} brought such prescriptions to 
a very simple form in two, three and four dimensions. 

In this work we focus on $D = 4$ dimensions and develop a Feynman diagrammatic method, 
improving significantly on Ref.~\cite{Baacke:2003uw}, and arrive at the same simple 
results of Refs.~\cite{Dunne:2005rt, Dunne:2006ct}, obtained with different methods 
for scalars.
Moreover, we extend our method to fermion and gauge boson fluctuations and
arrive at analogous simple and elegant prescriptions for the functional determinants 
of these species as well. 
To our knowledge these are new results. 
Fermions and gauge bosons are more involved and require a proper understanding of the group 
theoretical structure associated with the rotational invariance of the fluctuation operators. 
We work out these non trivial aspects in detail in the main body of the paper.

Even though there might be some tendency to dismiss the calculation of functional 
determinants and leave it to guesswork, we believe that times are mature to improve 
on the precision of vacuum decay rates' calculations, including an accurate determination
of the prefactor as well.
Refs.~\cite{Baacke:2003uw, Dunne:2005rt, Dunne:2006ct} already took important steps 
in that direction and a numerical package was developed~\cite{Ekstedt:2023sqc} where 
this kind of calculations are implemented. 
In light of this, the present work provides state-of-the-art prescriptions for 
functional determinants, which are completely generic and can be implemented in a 
large variety of models. 

The quantitative expression for the nucleation rate per unit volume $\gamma = \Gamma/V$ 
at one-loop is given by
\begin{equation}\label{eq:master}
\begin{split}
  \gamma=e^{-S} \,\left(\frac{S}{2\pi}\right)^2  {\rm Im}\sqrt{\left. \frac{{\det} 
  {\hat S}'' }
  {{\det}' S''} \right|_\phi} \,\left. \frac{{\det} S''}{{\det} {\hat S}'' } 
  \right|_\psi \, 
  V_G \,{J_G} \; \sqrt{\left. \frac{{\det} {\hat S}'' }{{\det}' S''} \right|_{A,a}} 
  \, \left. \frac{{\det} S''}{{\det} {\hat S}'' } \right|_{c \bar c} \, .
\end{split}
\end{equation}
Here, $S$ is the renormalised Euclidean bounce action and $S''$ is the second variational
derivative of $S$ with respect to various fields, evaluated on the bounce, whilst 
$\hat S''$ is evaluated on the FV.
We use $\hat{O}$ to label any operator $O$ or a function/constant that is evaluated 
on the FV throughout the paper.

The purpose of our work is to revisit and simplify, while keeping general validity,
the regularisation of the determinants in the scalar~$|_\phi$, fermion~$|_\psi$ 
and gauge/would-be Nambu-Goldstone boson $|_{A,a}$ sectors.

$S''|_{\phi}$ and $S''|_{A, a}$ are differential operators that may contain eigenmodes 
with zero eigenvalue, called zero modes.
These appear in the presence of symmetries and have to be removed from the determinant to 
get $\det '$.
The removal employs the use of collective coordinates~\cite{Gervais:1975yg, Callan:1977pt,
Coleman:1985rnk} and results in adding of various pre-factors to the rate: $(S/2\pi)^2$ 
for translations, $V_G$ for the volume of the gauge group $G$ and $J_G$ for the Jacobian of 
the gauge zero modes.
For the case of the Standard Model (SM) see e.g~\cite{Isidori:2001bm, Andreassen:2017rzq, 
Chigusa:2018uuj} and~\cite{Bhattacharya:2024chz} for zero removal subtleties.
To interpret $\gamma$ as a tunneling rate, the determinant needs to have an imaginary part. 
It comes from $S''|_\phi$, which has exactly one~\cite{Coleman:1987rm} negative eigenvalue, 
such that the square root in~\eqref{eq:master} becomes imaginary~\cite{Coleman:1977py}.
While important, these features are related to the infrared (IR) part of the spectrum. 
Instead, the focus of this work is on the ultraviolet (UV).

As usual in QFTs, calculating loop effects means dealing with UV divergences.
The functional determinants in spacetime dimensions $D\geq 2$ are no exception:
they are UV divergent and must be regularised. 
In this paper we provide simple prescriptions for such a regularisation in $D=4$.
Given the spherical $O(4)$ symmetry of the bounce~\cite{Coleman:1977th}, the various fluctuation
operators $S''$, whose determinant we need to compute, turn out to be rotationally invariant.
The eigenfunctions of $S''$ can then be decomposed into a radial and an angular part, employing
a generalisation of spherical harmonics, as $\sum_{\nu,m}c_{\nu,m}(\rho){\mathbb Y}_{\nu,m}(\hat \theta)$,
such that $S''$ acts diagonally in $\nu$ and $m$, respectively a total angular momentum quantum number
(whose precise definition for the fields of various spins is given later), and a polarisation index.
The determinant then factorises as ${\det} S'' = \prod_\nu ({\det} S''_\nu)^{d_\nu}$,
where $d_\nu$ is the degeneracy factor counting the independent polarisations at given $\nu$,
and ${\det} S''_\nu$ is a partial determinant.
Here, $S''_\nu$ is a differential operator in the radial variable $\rho^2 = t^2 + x_i^2$.
Using the Gelfand-Yaglom theorem~\cite{Gelfand:1959nq}, the product of eigenvalues of $S''_\nu$
is carried out.
It is the singular behaviour of the product 
$\prod_\nu \big(\det S''_\nu/\det \hat S''_\nu \big)^{d_\nu}$ 
at large $\nu$ that encodes the UV divergences.

One way of dealing with this divergent product was studied in Ref.~\cite{Baacke:2003uw} for 
the case of scalar fluctuations.
It amounts to isolating a few one-loop Feynman diagrams (on the bounce background), which are 
responsible for the UV divergences of the functional determinant. 
These diagrams are then evaluated in two manners on two different spaces.
The first uses a basis in coordinate space together with the multipole expansion (similar to 
${\det} S''$), so the variables at play are $\rho$ and $\nu$.
The second is done in momentum space with standard dimensionally-regulated integrals that produce
$1/\epsilon$ poles in combination with $\ln \mu$, with $\epsilon = 4 - D$.
The first expression is then subtracted from the determinant to make it convergent, while 
the second is added back to compensate for the subtraction. 
Schematically,
\begin{equation}\label{eq:procscheme}
  \ln \frac{\det S''}{\det \hat S''} = \sum_\nu \left( 
  d_\nu \ln\frac{\det S_\nu''}{\det \hat S_\nu''} - {\rm subtraction}_\nu\right) + 
  \left( {\rm addback} \right)_{\rm regularised} \, .
\end{equation}
One complication in these calculations is in the momentum space part, where one has to deal
with integrals of the form
\begin{equation} \label{eq:momdifficult}
  \int {\rm d}^D q \, {\tilde \phi}^2(q)\tilde{\phi}^2(-q) 
  \int \frac{\text{d}^D k}{(k^2 - m^2)((k+q)^2 - m^2)}  \, .
\end{equation}
Here $\tilde\phi(q)$ is the Fourier transform of the bounce field configuration $\bar\phi(\rho)$.
Typically $\tilde\phi(q)$ is a complicated function, the integral above cannot be solved 
analytically and one has to resort to numerical methods. 
A notable exception is the Fubini-Lipatov bounce~\cite{Fubini:1976jm, Lipatov:1976ny}, relevant for the calculation of the electroweak 
vacuum lifetime, for which the integral can be done analytically.

Ref.~\cite{Dunne:2005rt} revisited the problem and provided a simpler recipe using radial WKB 
and an angular momentum cutoff regularisation~\cite{Dunne:2004cp, Dunne:2004sx, Dunne:2005te}.
They also showed numerically that their result was in agreement with Ref.~\cite{Baacke:2003uw}. 
A year later, Ref.~\cite{Dunne:2006ct} used the $\zeta$-function regularisation method, with a 
derivation in generic $D$ dimensions, and arrived at a much simpler prescription, containing only
integrals in the coordinate $\rho$, no integrals in momentum space. 
They also showed that in $D=4$ their result agreed with those of 
Refs.~\cite{Baacke:2003uw, Dunne:2005rt} in the $\overline{\rm MS}$ scheme.

In this paper we adopt the dimensional regularization scheme, as in Ref.~\cite{Baacke:2003uw}.
Differently from them, we perform a double expansion of the Feynman diagram series, both in the 
coupling {\em and} in the squared mass parameter $m^2$, which appears in~\eqref{eq:momdifficult} 
and corresponds to the second derivative of the potential evaluated on the false vacuum. 
This is the first key to simplification. 
On top of it, we perform manipulations on the ``subtraction'' part to isolate the leading divergent 
terms in the $\nu$ series from the finite pieces. 
Last, in the ``addback'' part, we perform the inverse Fourier transform to revert back to 
coordinate space at the end of the calculation. 
To do so we take advantage of two identities which we derive in Appendix~\ref{app:FTProofs}.
When the dust settles, we obtain a very simple prescription corresponding to~\eqref{eq:procscheme} 
in terms of single integrals only in the coordinate $\rho$. 
Our result is identical to that of Ref.~\cite{Dunne:2006ct} for scalars, but obtained with a 
different method. 

Our method, which employs dimensional regularisation, not only reproduces the simple result for 
scalars, but is now extended to the analogous calculations involving fermions and gauge bosons.
Here, we perform the analysis also for these cases and manage to provide simple prescriptions
corresponding to~\eqref{eq:procscheme} in terms of single $\rho$-integrals only. 
Our results are general, in the sense that they work for a generic scalar potential and bounce, for 
a generic number of species of spin $0,1/2,1$, and easily generalise to non Abelian gauge bosons.

The final expression for the functional determinant is a regularised quantity, with divergences 
regulated by poles, with a residual dependence on the dimensionful parameter $\mu$, plus some
remaining finite parts.
The $1/\epsilon$ and $\ln\mu$ terms at the end of the day must cancel out against the analogous 
terms coming from the factor $e^{-S}$ in~\eqref{eq:master}, where $S$ is the renormalised bounce 
action at one-loop, in order to obtain the physical observable decay rate $\gamma$.

The remainder of the paper is organised as follows. 
In Sec.~\ref{sec:summary} we define the theoretical setup and give the prescriptions for obtaining 
the subtraction series and the compensating dimensionally-regulated term. 
After summarising the prescriptions, we give an explicit example of their use for the benchmark
computation of the SM lifetime. 
In Sections~\ref{sec:scalar},~\ref{sec:fermion} and \ref{sec:vector} we present a detailed 
derivation of our formul\ae\ for scalar, fermion and vector fluctuations. 
In Section~\ref{sec:outlook} we discuss possible interesting future directions where our results could apply.
A crucial step towards establishing our simple results are certain Fourier transform 
identities that are derived in Appendix~\ref{app:FTProofs} using two different techniques. 
In Appendix~\ref{app:analytical} we collect some analytic formul{\ae} relevant for the SM lifetime calculation.

%
%
\section{Summary} \label{sec:summary}

In this section we provide the essence of our findings, assuming some knowledge on how to set up 
the calculation of functional determinants~\cite{Callan:1977pt}.
We distil our results to compact formulas that can readily be used for any concrete problem.
We work in $D = 4$ Euclidean space-time dimensions 
and start with considering a weakly coupled theory for a set of real scalars 
$\phi^i$ and Dirac fermions $\psi^a$, and we generalize the procedure of dealing with Majorana 
fermions at the end of section~\ref{sec:fermion}.
We discuss separately the case of minimally coupled gauge bosons $A_\mu$. 
The most general Euclidean action for scalar and fermions can be expressed as
\begin{align} \label{eq:Lagrangian}
  S = \int_x \left( \frac{1}{2} \, \partial_\mu \phi^i\partial^\mu\phi^i+V(\phi^i) +
  i\bar{\psi}^a \slashed{\partial} \psi^a +\bar{\psi}^a {\bf{M}}_{ab}(\phi^i)\psi^b \right) \, ,
\end{align}
where $\slashed{\partial} = \gamma_\mu \partial^\mu$, with $\gamma_\mu$ the $4D$ Euclidean 
Dirac matrices.
Throughout the paper we use the shorthand notation for integrals,
\begin{align} \label{eq:shorthand}
 \int_x         &\equiv \int {\rm d}^4 x \, , &  
 \int_k         &\equiv \int \frac{{\rm d}^4 k}{\left( 2 \pi \right)^4} \, , & 
 \int_\rho      &\equiv \int_0^\infty {\rm d} \rho \, , &
 \int_\lambda   &\equiv \int_0^\infty {\rm d} \lambda \, .
\end{align}
Whereas for the spatial coordinates $x,y$ we only have $4D$ integrals, for the 
momentum coordinates $k,p,q$ we occasionally perform some in $D$ dimensions, in
which case we write them out explicitly; $\rho$ is the Euclidean radius, while 
$\lambda$ is the eigenvalue of the Laplacian operator in~\eqref{lameig}.
In~\eqref{eq:Lagrangian} the indices $i$ and $a, b$ denote flavor, 
and ${\bf M} = {\rm Re}{M} - i\gamma_5 {\rm Im} {M}$ is parametrised in terms of a single
complex matrix ${M}_{ab}(\phi^i)$ with ${\rm Re} M \equiv( M + M^\dagger)/2$ 
and ${\rm Im}M \equiv (M-M^\dagger)/(2 i)$, which covers scalar and pseudo-scalar couplings.
We assume that the model admits {\em one} $O(4)$ symmetric bounce solution 
$\bar{\phi}^i(\rho)$ which extremises the action and depends only on 
the Euclidean radius $\rho = \sqrt{x^\mu x_\mu}$ and goes to the FV constant $\hat{\phi}^i$ 
when $\rho \to \infty$. 
There is no loss of generality in choosing real scalars, since a complex scalar can
always be decomposed in two real fields.

The action can be expanded up to second order in the fluctuations around the bounce, with 
$\phi^i(x) = \bar\phi^i(\rho) + \delta\phi^i(x)$.
This gives $S = S_0 + S_2 + \ldots$, where $S_0 = S[\bar\phi]$ is the bounce action,
$S_1$ vanishes because the bounce is an extremum, and
\begin{align}\label{eq:S[2]}
  S_2 = \frac{1}{2} \int_x \, \left( \delta\phi^i\left(-\delta_{ij}\partial^2 +
  m^2_{ij}(\rho) \right)\delta\phi^j + 
  \bar{\psi}^a \left( i\delta_{ab} \slashed{\partial} + 
  {\bf M}_{ab}(\bar{\phi})\right)\psi^b \right) \, .
\end{align}
The operators $S''\vert_\phi$ and $S''\vert_\psi$ are obtained by taking the second 
variational derivative of $S_2$,
\begin{align}
  S''_{ij}\vert_\phi &= \frac{\delta^2 S_2}{
  \delta(\delta\phi^i) \delta(\delta\phi^j)} \, , 
  &
  S''_{ab}\vert_\psi &= \frac{\delta^2 S_2}{
  \delta\bar\psi^a \delta\psi^b} \, .
\end{align}

Given the rotational symmetry of the whole problem, the determinant ratios 
$\det S''/\det \hat S''$, with $\hat S''$ calculated on the FV, can be split into sectors 
with definite {\em total} angular momentum~\cite{Callan:1977pt}, which we label by $\nu$, 
an integer running from 1 to $\infty$. 
The determinant ratio is then a product of partial contributions $R_\nu$ for each $\nu$,  
with an additional degeneracy factor $d_\nu$, such that 
$\det S''/\det \hat S'' = \prod_\nu R_\nu^{d_\nu}$.
This infinite product is divergent and must be regularised. 
The procedure consists in isolating the UV divergent terms, subtracting them and adding 
back their regularised counterparts. 
In this work we adopt the dimensional regularisation scheme, which is familiar 
to particle physicists.

For scalar fluctuations we find
\begin{equation} \label{eq:DetS_scal}
\begin{split}
  \left.\ln \frac{{\det}' S''}{{\det} {\hat S}'' } \right|_\phi 
  =&  \sum_{\nu = 1}^\infty  \left( d_\nu^\phi \ln R_\nu^\phi - \frac{\nu}{2} \int_\rho \, 
  \rho \, \tr \delta m^2 + \frac{1}{8 \nu} \int_\rho \, \rho^3 \tr \delta m^4 \right)
  \\
  &-\frac{1}{8}  \int_\rho \, \rho^3 \tr\delta m^4 
  \left( \frac{1}{\epsilon} + 1+ \gamma_E + \ln \frac{\tilde \mu \rho}{2}  \right) \, ,
\end{split}
\end{equation} 
where we defined the field-dependent mass matrices and their constant FV values as
\begin{align}
  m^2_{ij}(\rho)    &\equiv \frac{\text{d}^2 V}{\text{d} \phi^i \text{d} \phi^j} 
  \bigg \vert_{\bar\phi} \, ,
  &
  \hat{m}^2_{ij}    &\equiv \frac{\text{d}^2 V}{\text{d} \phi^i \text{d} \phi^j}
  \bigg \vert_{\hat\phi}  \, ,
  \\ 
  \delta m^2_{ij} &\equiv m^2_{ij} - \hat{m}^2_{ij} \, ,
  &
  \delta m^4_{ij} &\equiv m^4_{ij} - \hat{m}^4_{ij} \, .
\end{align}

The trace goes over the flavor indices $ij$, while $d_\nu^\phi = \nu^2$ is the degeneracy
factor that counts the number of angular momentum modes with a given $\nu$. 
Such modes are the 4-dimensional generalisation of the spherical harmonics, the 
hyperspherical harmonics. 
Several comments about Eq.~\eqref{eq:DetS_scal} are in order.
\begin{itemize}
  \item[-] The prime at the numerator on the left hand side (LHS) means that one should 
  remove the zero eigenvalues from the determinant. 
  These are expected whenever the  bounce solution breaks some symmetry of the FV.
  This is always the case for Euclidean translations, because the bounce is an instanton,
  {\it i.e.} it is localised in (Euclidean) spacetime. 
  In practice this implies that $R_{\nu=2}^\phi$ requires a special treatment\footnote{We use 
  $\nu$ to label the angular momentum multipoles.
  In the literature various labels are used; e.g. comparing to Eq.~(3.14) in~\cite{Callan:1977pt}, 
  our $\nu$ corresponds to $2 j + 1$, while comparing to~\cite{Dunne:2006ct} $\nu = l + 1$.}, 
  explained in \cite{Callan:1977pt}. 
  Notice, however, that zero modes are characterised by a low $\nu$, while we are 
  interested in the limit $\nu \to \infty$. 
  Therefore, the special care required for dealing with zero modes has no impact on 
  the asymptotic subtraction that is our main focus.
  \item[-] Computing $R_\nu^\phi$ explicitly is difficult.
  It is often helpful to use the Gelfand-Yaglom method to make progress, see 
  {\it e.g.}~\cite{Dunne:2005rt,Dunne:2006ct}.
  Only in a few cases $R_\nu^\phi$ can be obtained analytically~\cite{Andreassen:2017rzq,
  Chigusa:2017dux, Ivanov:2022osf, Guada:2020ihz},
  but typically requires numerical evaluation~\cite{Isidori:2001bm, Matteini:2024xvg}.
  Here we are not concerned with any specific explicit form of $R_\nu^\phi$.
  We are interested in the remaining terms which universally regularise the expression.
  \item[-] The subtraction part in the first line of~\eqref{eq:DetS_scal} consists of 
  two terms that diverge quadratically and logarithmically at large $\nu$. 
  Diagrammatically, they are related to Feynman diagrams with one and two propagators, 
  as depicted in Fig.~\ref{fig:FD_FV}, but not exactly equal.
  They only keep the leading high-$\nu$ behaviour, which is needed to absorb the divergences 
  of the $d_\nu^\phi R_\nu$ series and make the sum over $\nu$ in the first line finite.
  In this sense this prescription is minimal and given in terms of single integrals 
  in coordinate rather than momentum space.
  \item[-] In the second line of~\eqref{eq:DetS_scal} we have the regulator 
  $\epsilon = 4 - D$ in the standard combination with $\ln \tilde \mu$, where 
  $\tilde \mu^2 = 4 \pi e^{-\gamma_E} \mu^2$ is the usual dimensionful parameter 
  introduced in dimensional regularisation.
  When the determinant ratio is combined with the renormalised action 
  following~\eqref{eq:master}, the $1/\epsilon$ and $\ln \tilde \mu$ terms must cancel out. 
\end{itemize}
A similar result to~\eqref{eq:DetS_scal} was obtained for scalars in~\cite{Dunne:2006ct}
with a $\zeta$-function regularisation.
In this paper we (re)derive this result for scalars using the more familiar techniques of
dimensional regularisation and generalise it to multiple scalars, fermions and vector bosons.

For fermionic fluctuations the result is
\begin{equation} \label{eq:DetS_ferm}
\begin{split}
  \left.\ln \frac{{\det} S''}{{\det} {\hat S}'' } \right|_\psi 
  =&  \sum_{\nu = 1}^\infty  \left( d_\nu^\psi \ln R_\nu^\psi -\left(\nu+\tfrac12\right)  
  \int_\rho \, \rho \, \tr \delta m_\psi^2 
 + \frac{1}{4 \nu} \int_\rho \, \rho^3 \tr \left( \delta m_\psi^4+{\dot{m}_\psi}^2 \right) \right)
 \\
 &  -\frac{1}{4}  \int_\rho \, \rho^3 \tr\left( \delta m_\psi^4+{\dot{m}_\psi}^2 \right) 
  \left( \frac{1}{\epsilon} + 1 + \gamma_E + \ln \frac{\tilde \mu \rho}{2}  \right) \, .
\end{split}
\end{equation}
Here $d_\nu^\psi = \nu(\nu + 1)$ is the degeneracy factor for fermions, and we defined
\begin{align}\label{eq:mpsidef}
  m^2_\psi(\rho) &\equiv \left. M^\dagger M \right|_{\bar{\phi}} \, , 
  &
  \hat{m}^2_\psi &\equiv \left. M^\dagger M \right|_{\hat{\phi}} \, , 
  &
  \dot{m}_\psi^2 &\equiv \left . \dot M^\dagger \dot M \right|_{\bar{\phi}} \, , 
  \\
  \delta m_\psi^2 &\equiv m_\psi^2 - \hat m_\psi^2 \, , 
  &
  \delta m_\psi^4 &\equiv m_\psi^4 - \hat m_\psi^4 \, , 
\end{align}
where $M_{ab}(\phi^i)$ is given below~\eqref{eq:shorthand}. 
The mass matrices are hermitian and are constructed from the $\rho$-dependent
$M(\bar{\phi}(\rho))$ and its $\rho \to \infty$ FV limit $M(\hat{\phi})$, while
the dot stands for the derivative over $\rho$, i.e. $\dot M = \partial_\rho M$.

There are some differences with respect to the scalar case. 
The first is the appearance of $\dot{m}_\psi$, related to the fact that 
we are dealing with massive Dirac fermions that couple two Weyl components.
The second is the form of the asymptotic subtraction, which has a modified $\nu$
dependence in the term proportional to $\delta m_\psi^2$.
Finally, the degeneracy factor $d_\nu^\psi$ is different from the scalar case, which 
follows from the fact that Dirac fermions belong to the $(0,1/2) \oplus (1/2, 0)$ 
representation of the Lorentz group.
Despite these differences, the similarity to~\eqref{eq:DetS_scal} is remarkable:
for fermions we also have a quadratic and a logarithmic UV divergence, 
and the regularization prescription only involves single integrals in $\rho$.

For gauge boson fluctuations, we consider a simple setup with a $U(1)$ gauge field 
$A_\mu$ and a charged scalar field $\Phi$, and we comment below on the non-Abelian 
generalisation.
The Euclidean action is given by
\begin{equation} \label{eq:Lgauge}
  S[\Phi,A] = \int_x \left( \frac{1}{4} F_{\mu\nu}^2 + \left| D_\mu \Phi \right|^2 + 
  V(\Phi) + L_{\rm GF} + L_{\rm ghost} \right),
\end{equation}
where $L_{\rm GF}$ and $L_{\rm ghost}$ are the gauge fixing term and the Faddeev-Popov 
(FP) ghost term, and the gauge coupling $g$ is defined through
$D_\mu \Phi = (\partial_\mu - i g A_\mu) \Phi$.
The scalar field is decomposed as $\Phi = (\phi + i a)/\sqrt 2$, where $\phi$ is the component
that carries the FV VEV $\hat \phi$ and/or a non-trivial bounce profile $\bar \phi$. 
We adopt the following gauge fixing,
\begin{align} \label{eq:Lghost}
  L_{\rm GF}^{R_\xi} &= \frac{1}{2 \xi}\left(\partial_\mu A_\mu- g \bar\phi(\rho) a\right)^2 \, ,
  &
  L_{\rm ghost} &= \bar c\big({-}\partial^2 +g^2 \bar\phi(\rho)^2\big)c\, .
\end{align}
This is the background $R_\xi$-gauge, and it is convenient for our purposes to
fix $\xi = 1$, as was done in~\cite{Isidori:2001bm, Branchina:2014rva}. 
We expand the action~\eqref{eq:Lgauge} to quadratic order in fluctuations about 
the bounce solution,
\begin{equation}
\begin{split}
  S_2=&\int_x \left[\frac{1}{2}A_\mu \left(-\partial^2 + m_A^2(\rho) \right)\delta_{\mu\nu} 
  A_\nu + \frac{1}{2}a \left(-\partial^2 + m_a^2(\rho) \right)a \right. 
  \\ 
  & \left. \qquad + 2 A_\mu \left( g \partial_\mu \bar\phi \right) a
  + \frac{1}{2} \delta\phi \left( - \partial^2+m^2(\rho) \right) \delta\phi
  + \bar c \left(-\partial^2 + m_A^2(\rho) \right) c \right] \, .
\end{split}
\end{equation}
Here $m_A^2(\rho)$ and $m_a^2(\rho)$ are given in~\eqref{eq:maA}, while 
$m^2(\rho) \equiv \frac{{\rm d}^2}{{\rm d}(\delta\phi)^2}\left.
V\left(\frac{\bar\phi + \delta\phi}{\sqrt{2}} \right)\right\vert_{\delta\phi=0}$.
We see that $\delta\phi$ and $c\bar c$ are decoupled from each other and from the other fields; 
their contribution to~\eqref{eq:master} is formally identical to that of scalars. 
Here we focus on the remaining five degrees of freedom $a$ and $A_\mu$, which couple to 
each other.
The vector field $A_\mu$ requires care when expanded in total angular momentum $\nu$-modes, 
as it is not equivalent to a collection of four scalars; this was elaborated upon
in~\cite{Baratella:2024hju}. 
Our result for gauge boson fluctuations is
\begin{equation}\label{eq:DetS_vect}
\begin{split}
  \left. \ln \frac{\det'S''}{\det\hat S''}\right|_{A,a}^{\xi=1} &= 
  \sum_{\nu=1}^\infty \Bigg( d_\nu^\phi \ln R_\nu^{(Da)} + 
  2 \, d_\nu^T \ln R_\nu^{(T)} 
  \\
  &- \frac12 \int_\rho \rho \left(\nu\,\delta m_a^2 +
  2(2 \nu + 1 )\, \delta m_A^2 \right)
  +\frac1{8 \nu} \int_\rho \rho^3 \left(\delta m_a^4 + 4 \, 
  \delta m_A^4 + 8 \, \dot m_A^2\right) \Bigg) 
  \\
  & +\frac{1}{2}\int_\rho \rho\,\delta m_A^2+\frac{1}{8}\int_\rho \rho^3\delta m_A^4 
  \\
  & 
  -\frac18 \int_\rho \rho^3\left(\frac1\epsilon + 1 + \gamma_E +
 \ln \frac{\tilde \mu \rho}2 \right) \left(\delta m_a^4 + 4 \, \delta m_A^4 +
 8 \, \dot m_A^2 \right) \, .
\end{split}
\end{equation}
Here we defined the field-dependent masses
\begin{align} \label{eq:maA}
  m_A^2(\rho)       &\equiv g^2 \, \bar \phi^2(\rho) \, , &
  m_a^2(\rho)       &\equiv g^2 \, \bar \phi^2(\rho) + V_{aa} \left( \bar \phi \right) \, , &
  \dot m_A(\rho)    & \equiv g \, \partial_\rho \bar\phi(\rho) \, ,  
  \\
  V_{aa}(\bar\phi)  &\equiv \tfrac{{\rm d}^2}{{\rm d}a^2}V
  \left( \tfrac{\bar\phi+i a}{\sqrt 2} \right) \big|_{a = 0} \, , &
  \delta m^2_{a,A}  &\equiv m^2_{a,A}-\hat m_{a,A}^2 \, , &
  \delta m^4_{a,A}  &\equiv m^4_{a,A}-\hat m_{a,A}^4 \, ,  
\end{align}
and their FV values $\hat m_{a,A}^2 = \lim_{\rho \to \infty} m_{a,A}^2(\rho)$.
\begin{itemize}
  \item[-] The prime at the numerator indicates removal of zero modes, 
  which has to do with the lowest multipoles, as in the scalar case. 
  We refer the interested reader to Appendix B in~\cite{Chigusa:2018uuj} or Section 5.3 
  in~\cite{Andreassen:2017rzq} for the zero removal procedure in the gauge sector. 
  \item[-] The partial determinants split into two sectors: one contains three degrees of 
  freedom corresponding to the two `diagonal' $D$ modes of $A_\mu$, which mix with the would 
  be NG boson $a$;
  the other contains the two transverse $T$ modes of $A_\mu$. 
  The objects in the former have multiplicity $d_\nu^\phi = \nu^2$ as for the scalars, but
  those in the latter have $d_\nu^T = \nu(\nu + 2)$~\cite{Baratella:2024hju}.
  This distinction is important for arriving at the final simple result in~\eqref{eq:DetS_vect},
  which would not simplify with the wrong degeneracy factor.
  We are not interested in explicit expressions for $R_\nu^{(Da)}$ and $R_\nu^{(T)}$, but 
  only in the remaining terms that regularize the ratio of functional determinants. 
  \item[-] We presented a common subtraction term for the $Da$ and $T$ sectors. If one is 
  interested in separately regulating the two, the splitting is 
  \begin{equation}
  \begin{split}
    2(2\nu+1)\delta m_A^2 & =\left[(2\nu+\tfrac 2\nu)_{\phi}+
    (2\nu+2-\tfrac 2\nu)_T\right]\delta m_A^2 \, , 
    \\
    4\,\delta m_A^4& =(2_\phi+2_T)\,\delta m_A^4 \, ,
  \end{split}
  \end{equation}
  in the second line of \eqref{eq:DetS_vect}.
  The suffixes $\phi,T$ indicate which part should be subtracted from $d_\nu^{\phi,T}\ln R_\nu$. 
  All the other terms, proportional to $\delta m_a^{2,4}$ or $\dot m_A^2$, belong to the $Da$ sector.
  \item[-] The subtraction series and compensating term are expressed, as before, out 
  of single integrals of $\rho$-dependent masses and their derivatives.
  Similarly to the fermionic sector, the appearance of $\dot m_A$ is rooted in the 
  presence of an off-diagonal term in the fluctuation matrix that mixes the would-be 
  NGB $a$ with the $D$-polarisations of $A_\mu$, see Section \ref{sec:vector}.
  \item[-] Physically, it makes sense to separate the cases in which the FV is 
  $U(1)$-symmetric from those in which it is not; our formula is universal, in that 
  it covers both situations. 
  Then it is either $g^2 \hat \phi^2 = 0$ or $V_{aa}(\hat\phi) = 0$.
  \item[-] Comparing against scalars~\eqref{eq:DetS_scal} and fermions~\eqref{eq:DetS_ferm}, 
  we note the appearance in the third line of~\eqref{eq:DetS_vect} of two extra terms,
  \begin{equation}\label{eq:non_uni}
     \frac{1}{2}\int_\rho \rho\,\delta m_A^2+\frac{1}{8}\int_\rho \rho^3\delta m_A^4\,.
  \end{equation}
  For the origin of the first one, see comment below~\eqref{eq:T10vect} in Sec.~\ref{sec:vector}.
  The second one comes from the fact that, with the standard dimensional continuation 
  rules we adopt, the vector has $4{-}\epsilon$ degrees of freedom instead of $4$, 
  leaving this $\epsilon/\epsilon$ finite remnant (at variance with scalars and fermions, 
  that maintain the same number of degrees of freedom when dimensionally continued).
  \item[-] The object in \eqref{eq:DetS_vect} is gauge-dependent. 
  However, physical quantities like $\gamma$ \eqref{eq:master} must not be. 
  It was proven in~\cite{Endo:2017gal} that the combination
  \begin{equation} \label{eq:gauge_inv}
    \left. \frac{\det'S''}{\det\hat S''}\right|_{A,a}\left( \left.\frac{\det S''}{
    \det \hat S''}\right|_{c\bar c}\right)^{-2} \, ,
  \end{equation}
  is gauge invariant.
  In order to compare results in different gauges, one has to take the ghosts' 
  contribution into account.
  \item[-] The result was derived for an abelian gauge theory, but it can be
  straightforwardly extended to the non-abelian case. 
  Working to quadratic order in the fluctuations, the non-abelian nature of the group 
  does not show up, as the quadratic action naturally splits into sectors that are 
  $U(1)$-like.
  The only trace of the SM non-abelian structure in~\eqref{eq:master} is in the factor 
  $V_G$, that accounts for the equivalent directions in the Higgs doublet space, in which 
  the bounce could nucleate, and is given by $V_G={\rm Vol}(S^3)=2\pi^2$, 
  see~\cite{Andreassen:2017rzq,Chigusa:2018uuj} for more details. 
\end{itemize}
In the next section we show how to use~\eqref{eq:DetS_vect} explicitly in the SM calculation.

%
%
\subsection{Standard Model} \label{sec:SMcheck}

Our purpose in this section is twofold. 
First, we want to exemplify the use of our prescriptions on a concrete and important example,
showing that our results are ready-to-use and simple.
Second, we want to check our formul{\ae}  against SM results in previous literature.

\vspace{.3cm}
\textbf{Physical Higgs boson.} 
We start with the contribution to the functional determinant coming from the fluctuations of 
the physical Higgs boson $h$, following the notation of~\cite{Baratella:2024hju}.
The scalar potential is given by $V(h)=\frac{1}{4}\lambda h^4$, with $\lambda<0$. 
The metastable FV solution corresponds to $\hat{h}=0$, while the bounce is given by the 
Fubini-Lipatov~\cite{Fubini:1976jm, Lipatov:1976ny} solution
\begin{equation} \label{eq:h_bounce}
  \bar{h}(\rho) = \sqrt{\frac{8}{-\lambda}}\frac{R}{R^2+\rho^2} \, ,
\end{equation}
with $R$ an arbitrary parameter that sets the instanton size.
We need a simplified version of~\eqref{eq:DetS_scal} with a single flavor $i$, so we 
can drop the `tr'. 
The FV mass is trivial $\hat{m}_h^2=0$, while 
$m_h^2(\rho) = 3\lambda \bar{h}^2(\rho) = -24 R^2/(R^2+\rho^2)^2$. 
We only need to compute the two simple integrals
\begin{align}
  \int_\rho \, \rho \,      m_h^2 &= -12 \, ,  &
  \int_\rho \, \rho^3 \,    m_h^4 \left( a + b \ln \rho \right) & = 
  48 \left( a + b \ln R \right) \, ,
\end{align}
and plug them into~\eqref{eq:DetS_scal} to get
\begin{equation}\label{eq:DetS_h}
  \left.\ln \frac{{\det}' S''}{{\det} {\hat S}'' } \right|_h 
  = \sum_{\nu = 1}^\infty \left( d_\nu^h \ln R_\nu^h+{6}{\nu} + \frac{6}{\nu} \right)  
  - 6 \left( \frac{1}{\epsilon} + 1 + \gamma_E+\ln \frac{\tilde{\mu} R}{2} \right) \, .
\end{equation}
To evaluate the sum we need the partial determinant 
$R_\nu^h=(\nu-1)(\nu-2)/(\nu+1)/(\nu+2)$, which can be found for example 
in~\cite{Andreassen:2017rzq, Chigusa:2018uuj}. 
At large $\nu$ we have $\ln R_\nu^h=-6/\nu - 6/\nu^3 + O(\nu^{-4})$, so the sum
in~\eqref{eq:DetS_h} is convergent.
This is already a first non-trivial check of our subtraction prescription. 
To carry out the sum explicitly there is another issue to take care of. 
For $\nu=1,2$ the partial determinant $R_\nu^h$ vanishes due to the presence of zero 
modes, so it has to be cured.
We defer to previous literature for a discussion of this delicate 
point~\cite{Andreassen:2017rzq,Chigusa:2018uuj},
noting again that this problem has to do with the low-$\nu$ end of the multipole expansion, 
while the results of the present work deal with the high-$\nu$ end. 
Following~\cite{Andreassen:2017rzq}, one can consider the operator 
$S''_{\rm red}\equiv m_h^{-2}(\rho) S''$. 
The advantage of this choice is that the usual zero mode removal procedure works 
both for $\nu=1$, corresponding to scale invariance, and for $\nu=2$, corresponding to 
translational invariance. 
Moreover, the non singular determinant ratios ($\nu>2$)  and the subtraction terms are the 
same as for $S''$. 
After zero-removal~\cite{Andreassen:2017rzq} one finds $R_1^h=-1/5$ and $R_2^h=1/10$. 
With these values we can perform the sum in~\eqref{eq:DetS_h} and find
\begin{equation}\label{eq:h_explicit}
  \left. \ln \frac{{\det}' S''_{\rm red}}{{\det} {\hat S}''_{\rm red} } \right|_h =
  \ln(-1)-\frac{6}{\epsilon} - 6\ln\frac{\tilde{\mu}R}{2}+\frac{5}{2}-
  5\ln\frac{5}{6}-12\zeta'(-1) \, .
\end{equation}
The result agrees with~\cite{Andreassen:2017rzq} and reaffirms the validity 
of~\eqref{eq:DetS_scal}. 
The appearance of $\ln(-1)$ is crucial, because when we exponentiate~\eqref{eq:DetS_h} 
and take the reciprocal of its square root as dictated by~\eqref{eq:master}, it gives 
an imaginary part to the vacuum to vacuum transition amplitude, implying instability, 
i.e. the presence of a nonzero decay rate.

\vspace{.3cm}

\textbf{Top quark.} 
Following the same spirit we adopted for the physical Higgs fluctuations, we now consider 
the effect of top quark fluctuations on the SM decay rate. 
Using the standard normalisation for the Yukawa coupling, we have simply 
${\bf M}(h) = y_t h/\sqrt{2}$, for each of the $N_c=3$ color components. 
We find again that $\hat{m}_t = 0$, while 
$m_t^2(\rho)= y_t^2 \bar{h}^2/2 = 4 y_t^2/|\lambda| \, R^2/(R^2+\rho^2)^2$ 
and $\dot{m}_t(\rho)=-4\,y_t /\sqrt{|\lambda|} \, \rho R/(R^2+\rho^2)^2$. 
The relevant integrals are
\begin{equation}
\begin{split}\label{eq:top_integrals}
  \int_\rho \, \rho\, {\rm tr} \, m_t^2 &= 2\, N_c \, x_t \, , 
  \\
  \int_\rho \, \rho^3\, {\rm tr}\, \left( m_t^4 + \dot{m}_t^2 \right)
  &= \frac{4}{3} N_c \left( 2 x_t + x_t^2 \right) \, , 
  \\
  \int_\rho \, \rho^3 \ln \rho\, {\rm tr}\, \left( m_t^4 + \dot{m}_t^2 \right) &=
  \frac{4}{3} \, N_c \left(2 x_t + x_t^2 \right) \ln R + 2 \, N_c \, x_t \, ,
\end{split}
\end{equation}
where we set $x_t = y_t^2/(-\lambda)$ to avoid clutter and take the trace over color indices
to get $N_c$. 
The partial determinants are given in \cite{Andreassen:2017rzq, Chigusa:2018uuj} and read
\begin{equation}
  R_\nu^t = \bigg(\frac{\Gamma(\nu{+}1)^2}{\Gamma\!\left(\nu{+}1{+}i\sqrt{x_t}\right)
  \Gamma\!\left(\nu{+}1{-}i\sqrt{x_t}\right)}\bigg)^{\!\!2 N_c}
  \simeq \exp\left[2N_c\left(\frac{x_t}{\nu}-\frac{x_t}{2\nu^2} - 
  \frac{x_t^2{-}x_t}{6\nu^3}\right) \right] \, ,
\end{equation}
up to $O(\nu^{-4})$ terms at the exponent.
It is straightforward to check that our subtraction part from~\eqref{eq:DetS_ferm}, which 
equals $(\nu+\frac12)2N_cx_t-\frac{N_c}{3\nu}(x_t^2{+}2x_t)$, has precisely the same 
asymptotic behaviour for large $\nu$ as $d_\nu^t\ln R_\nu^t$, down to $O(\nu^{-1})$. 
Plugging the integrals from~\eqref{eq:top_integrals} into~\eqref{eq:DetS_ferm}, we find 
(omitting the overall color factor)
\begin{align}\label{eq:psi_ours} \nonumber 
    \ln \frac{\det S''}{\det \hat S''}\bigg|_t
    &=\sum_{\nu=1}^\infty \bigg(2 \nu(\nu{+}1) \ln \
    \frac{\Gamma(\nu+1)^2}{\Gamma\!\left(\nu{+}1{+}i\sqrt{x_t}\right)\Gamma\!\left(\nu{+}1{-}
    i\sqrt{x_t}\right)}-(2\nu{+}1)x_t+\frac 1 {3\nu}(x_t^2{+}2x_t)\bigg) 
    \\
    &-\frac13(x_t^2+2x_t) \bigg( \frac{1}{\epsilon}+1+\gamma_E+\ln\frac{\tilde\mu R}{2}\bigg)-\frac{x_t}{2}  
    \\ \nonumber
    &=\Sigma_\psi(x_t)-\frac13(x_t^2+2x_t) \bigg( \frac{1}{\epsilon}+1+\gamma_E+
    \ln\frac{\tilde\mu R}{2}\bigg)-\frac{x_t}{2} \,,
\end{align}
in agreement with~\cite{Andreassen:2017rzq, Chigusa:2018uuj}. 
The analytic expression for the sum of the $\nu$ series in the first line 
of~\eqref{eq:psi_ours}, dubbed $\Sigma_\psi(x_t)$, is given in Appendix~\ref{app:analytical}.

\vspace{.3cm}
\textbf{\textit{Z} and \textit{W} bosons.} 
As a final instructive example we apply Eq.~\eqref{eq:DetS_vect} to the SM gauge sector. 
We evaluate the determinant ratio for $Z$ boson fluctuations; the extension to 
$W$s simply amounts to changing $g_Z\to g_W$ and multiplying by 2, 
since $W$ comes with charge $\pm 1$. 
In the FV (limit of $v \to 0$) we have $\hat m_Z = \hat m_z=0$, while 
$m_Z^2 = g_Z^2 \bar{h}^2/4$, $m_z^2=(\lambda + g_Z^2/4) \bar h^2$ and 
$\dot m_Z = g_Z \, \partial_\rho \bar h/2$, with $\bar h$ given in~\eqref{eq:h_bounce}.

To test our formalism against previous literature on the SM lifetime some caution is required.
In Refs.~\cite{Andreassen:2017rzq, Chigusa:2018uuj}, on which we rely for part of our test, 
the wrong degeneracy factor for the transverse modes was used, see~\cite{Baratella:2024hju}. 
We correct that point here as well and furthermore take into account that parts of the calculations 
are done in different gauges.
In this work we use the $R_\xi$ gauge background gauge, see~\eqref{eq:Lghost}. 
Refs.~\cite{Andreassen:2017rzq,Chigusa:2018uuj} worked (mostly) in 
the Fermi gauge, defined by $L_{\rm GF}^{\rm Fermi} = \frac{1}{2}(\partial_\mu A_\mu)^2$ in 
place of $L_{\rm GF}^{R_\xi}$.
To make the comparison we follow the observation below~\eqref{eq:gauge_inv}. 
We start with $R_\nu^{(Da)}$ and $R_\nu^{(T)}$ in Fermi gauge from~\cite{Chigusa:2018uuj}.
There is no ghost contribution to~\eqref{eq:gauge_inv} in this gauge, the corresponding 
functional determinant is 1. 
Since the combination of $\{A_\mu,a,c,\bar c\}$ is gauge invariant, we then define the
subtraction prescription as in~\eqref{eq:DetS_vect}, safe we also subtract the ghost 
contribution in the same $R_\xi$ gauge, with $\xi = 1$. 
To subtract the ghost contribution, we need to recall that it is a complex scalar 
with fermionic statistics. 
Its determinant is regulated as in~\eqref{eq:DetS_scal}, with $\delta m^2(\rho)=m_Z^2$ 
(valid for $\xi = 1$), an overall factor of 2 because the scalar is complex, and a minus 
sign due to fermionic statistics.
In the end we must add back the dimensionally regulated version of what we subtracted.
Schematically, this amounts to
\begin{equation}\label{eq:vector_check}
\begin{split}
  & \ln \left. J_G^{-2}\,\frac{\det'S''}{\det\hat S''}\right|_{Z,z}-2
  \ln \left.\frac{\det S''}{\det\hat S''}\right|_{c\bar c} 
  \\
  &=\sum_\nu \bigg( d_\nu^\phi \ln R_\nu^{(Da)}|_{\rm Fermi} +  2 \, d_\nu^T 
  \ln R_\nu^{(T)}|_{\rm Fermi} + {\rm sub}_\nu\left[\eqref{eq:DetS_vect}\right]_{R_\xi} -2\,
  {\rm sub}_\nu\left[\eqref{eq:DetS_scal}\right]_{\delta m^2\to m_Z^2}\bigg) 
  \\ 
  &+{\rm addback}\left[\eqref{eq:DetS_vect}\right]_{R_\xi} 
  -2 \ {\rm addback}\left[\eqref{eq:DetS_scal}\right]_{\delta m^2\to m_Z^2}  \,.
  \end{split}
\end{equation}
The factor of $J_G^{-2}$ comes about because the gauge zero modes are not normalisable in the
presence of a scale-invariant bounce. 
This implies that $\det'$ is zero and only when we combine it with $J_G$, which is also 
singular, we get a finite result~\cite{Chigusa:2018uuj}
\begin{equation}
  R_1^{(Da)}|_{\rm Fermi}\equiv J^{-2}_G \left(\left.
  \frac{\det'S''}{\det\hat S''}\right|_{Z,z}\right)_{\!\!\nu=1,\,{\rm Fermi}}
  =\frac{|\lambda|}{16\pi} \, .
\end{equation}
Then we have~\cite{Andreassen:2017rzq, Chigusa:2018uuj}
\begin{align}
    R_{\nu\geq 2}^{(D,a)}|_{\rm Fermi}&=\frac{\nu-1}{\nu+1}\frac{\Gamma(\nu)\Gamma(\nu{+}1)}{
    \Gamma\!\left(\nu{+}\frac{1}{2}(1-\sqrt{1-8y_z})\right)\Gamma\!
    \left(\nu{+}1{-}\frac{1}{2}(1-\sqrt{1-8y_z})\right)} \, , 
    \\
    R_\nu^{T}|_{\rm Fermi}&=\frac{\Gamma(\nu{+}1)\Gamma(\nu{+}2)}{\Gamma\!
    \left(\nu{+}1{+}\frac{1}{2}(1-\sqrt{1-8y_z})\right)\Gamma\!
    \left(\nu{+}2{-}\frac{1}{2}(1-\sqrt{1-8y_z})\right)} \, ,
\end{align}
where $y_z = g_Z^2/(4|\lambda|)$. 
The subtraction terms are easy to evaluate
\begin{equation}
\begin{split}
  {\rm sub}_\nu\left[\eqref{eq:DetS_vect}\right]_{R_\xi} & =
  - \frac12 \int_\rho \rho \left(\nu\,m_z^2 + 2(2 \nu + 1 )\, m_Z^2 \right)
  +\frac1{8 \nu} \int_\rho \rho^3 \left( m_z^4 + 4 \, m_Z^4 + 8 \, \dot m_Z^2\right) 
  \\
  &= 2 \left( \nu - y_z \left( 5\nu + 2 \right) \right) +
  \frac{2}{3\nu} \left(1 + 5 y_z \right) \left(1 + y_z \right) \, ,
  \end{split}
\end{equation}
while for the ghost contribution we have
\begin{equation}
  {\rm sub}_\nu\left[\eqref{eq:DetS_scal}\right]_{\delta m^2\to m_Z^2} = 
  -\frac{\nu}{2}\int_\rho \rho \, m_Z^2 + \frac{1}{8\nu}
  \int_\rho \rho^3 \, m_Z^4 = -2 y_z\nu + \frac{2y_z^2}{3\nu}\,.
\end{equation}
The first non-trivial check on our prescription is that the sum $\sum_\nu$
in~\eqref{eq:vector_check} must be convergent. 
Indeed we find an exact cancellation of the divergent terms; the summand reduces to 
$4\nu^{-2}(y_z + y_z^2/3)+O(\nu^{-3})$. 
The remaining step is to compute the 'add-back' pieces and combine everything
\begin{equation}
\begin{split}
  & {\rm addback}\left[\eqref{eq:DetS_vect}\right]_{R_\xi} 
  -2 \ {\rm addback}\left[\eqref{eq:DetS_scal}\right]_{\delta m^2\to m_Z^2}
  \\
  &=\frac{1}{2}\int_\rho \rho \ m_Z^2 + \frac{1}{8} \int_\rho \rho^3 \ m_Z^4 -
  \frac{1}{8}\int_\rho \rho^3 \left(\frac1\epsilon+1+\gamma_E+\ln\frac{\tilde\mu\rho}2\right)
  \left(m_z^4+2\,m_Z^4+8\,\dot m_Z^2\right)
  \\
  &=2y_z+\frac{2}{3} y_z^2-\frac23(1+6y_z+3y_z^2)\left(\frac1\epsilon+1+\gamma_E+
  \ln\frac{\tilde\mu R}2\right)-4y_z\,.
\end{split}  
\end{equation}
The ghost contribution, calculated from~\eqref{eq:DetS_scal}, changes the term $4m_Z^4$ 
in the last integral in the second line, coming from \eqref{eq:DetS_vect}, into $2m_Z^4$.
When we put all the pieces together, we find
\begin{align}\label{eq:vector_SM}
  \ln J_G^{-2} \, &\frac{\det'S''}{\det\hat S''}\Bigg|_{Z,z}-2\ln \left.
  \frac{\det S''}{\det\hat S''}\right|_{c\bar c}
  \\
  &= \Sigma_{U(1)}(y_z)-\frac23(1+6y_z+3y_z^2)\left(\frac1\epsilon+1+\gamma_E+
  \ln\frac{\tilde\mu R}2\right)-2y_z+\frac{2}{3} y_z^2\, , \nonumber
\end{align}
where the sum $\Sigma_{U(1)}(y_z)$ admits an analytic expression, which is given in 
Appendix~\ref{app:analytical}. 
This final result for the regularised functional determinant in the SM gauge sector is an 
update and a revision of the results in~\cite{Andreassen:2017rzq, Chigusa:2018uuj}, taking 
into account the correct degeneracy factor $d_\nu^T$ for the transverse 
modes in~Ref.~\cite{Baratella:2024hju}.

{\bf Cancellation of divergences.} 
Another check on our prescription against the known SM results is to consider the cancellation
of poles and the renormalization scale dependence.
Let us consider the functional determinants in~\eqref{eq:master} with $\phi = h$, the physical 
Higgs fluctuation, $\psi = t$, the top quark, $(A,a) = (Z,W,a_Z, a_W)$, and $c\bar c$ their
corresponding ghosts.
Using~\eqref{eq:DetS_scal},~\eqref{eq:DetS_ferm},~\eqref{eq:DetS_vect}, and isolating the 
$1/\epsilon$ terms, we have
\begin{equation} \label{expeps1}
  {\rm exp} \left[\frac 1 \epsilon \frac{1}{16}\int_\rho \rho^3 \left( 
  m_h^4 {-} 4 N_c \left( m_t^4 {+} \dot m_t^2 \right) {+} \left( m_z^4 {+} 2 m_Z^4 {+} 8
  \dot m_Z^2 \right) {+} 2 \left(m_w^4 {+} 2m_W^4 {+} 8 \dot m_W^2 \right)\right) \right] .
\end{equation}
The integrals are readily evaluated using
\begin{align}
 \int_\rho \rho^3\bar h^4(\rho)         &= \frac{16}{3\lambda^2} \, , 
 &
 \int_\rho \rho^3\dot{\bar h}^2(\rho)   &=-\frac{16}{3\lambda} \, ,
\end{align}
which gives
\begin{equation}
  {\rm exp} \left[ \frac{2}{\epsilon}
  \left( 2 - \frac{g_Z^2+2g_W^2}{4\lambda} + \frac{g_Z^4+2g_W^4}{32\lambda^2} + 
  \frac{N_cy_t^2}{3\lambda} - \frac{N_cy_t^4}{6\lambda^2} \right) \right] \, ,
\end{equation}
and perfectly agrees\footnote{
We have $2/\epsilon$ instead of $1/\epsilon$, due to our convention $D = 4-\epsilon$ as 
opposed to their $D = 4 - 2\epsilon$. 
To compare our result to theirs, one further needs $g^2 = g_W^2$, and 
$g^2 + (g^\prime)^2 = g_Z^2$.} 
with eq.~(6.10) in~\cite{Andreassen:2017rzq}.
In turn, this piece from the functional determinant exactly cancels against the renormalized 
tree-level action on the bounce, see eq.~(6.11) in~\cite{Andreassen:2017rzq}. 
This is a crucial point and an important check. 
The $1/\epsilon$ terms always come together with $\ln \mu$ in the same combination 
in~\eqref{eq:DetS_scal},~\eqref{eq:DetS_ferm} and~\eqref{eq:DetS_vect}. 
This example shows that when one assembles the result to compute the physical observable 
$\gamma$, the $1/\epsilon + \ln \mu$ terms cancel out.

%
%
\section{Scalars}\label{sec:scalar}

In this section we derive the formula~\eqref{eq:DetS_scal} for the scalar sector determinant 
ratio of the fluctuation operators over the bounce and FV configurations, which is part 
of the prefactor appearing in the false vacuum decay rate formula, Eq.~\eqref{eq:master}. 
In the next sections we are going to provide analogous results for the fermionic and 
gauge-bosonic sectors.

Let us consider here the simplest case of a single real scalar $\phi(x)$ 
with a potential $V$, a false vacuum constant solution $\hat{\phi}$ and 
a radially symmetric bounce $\bar\phi(\rho)$. 
The inclusion of multiple fields $\phi^i(x)$ with an arbitrary potential 
is postponed to the end of the section.
Starting from the Euclidean action
$S[\phi] = \int_x \left( \frac{1}{2}(\partial_\mu \phi)^2 + V(\phi) \right)$,
the operators obtained from the second variation of the action, {\it i.e.} the scalar 
fluctuation operators, are given by
\begin{align} \label{scalarSppdef}
  S^{\prime \prime}         &= -\partial^2 + m^2(\rho) \, , 
  &
  {\hat S}^{\prime \prime}  &= -\partial^2 + \hat m^2 \, ,
\end{align}
where the field dependent mass is given by
$m^2(\rho) = \frac{{\rm d}^2 V}{{\rm d} \phi^2} \vert_{\bar\phi}$
on the bounce and the constant
$\hat m^2 = \frac{{\rm d}^2 V}{{\rm d} \phi^2} \vert_{\hat \phi}$ is the
usual FV mass of the scalar. 
They act on the scalar fluctuations $\delta\phi(x)$ and are radially 
symmetric operators. 
This is because the bounce $\bar \phi(\rho)$ is a function of $\rho$ only,
and this dependence goes into the second derivative of the potential $m^2(\rho)$.
On the false vacuum $\hat \phi$, the second derivative is a constant $\hat m^2$.

The operator $S^{\prime \prime}$ always contains a single negative eigenvalue and 
$4$ vanishing eigenvalues because of translation symmetry (or more, in case of 
additional symmetries). 
The latter must be removed from the determinant and usually a prime is included 
to indicate the zero removal.
For the present discussion the well understood complications coming from 
zero eigenvectors or `zero modes' are irrelevant and we omit the prime.

\vspace{.3cm}
{\bf Multipole decomposition.} 
Because $m^2(\rho)$ is a spherically symmetric function of $\rho$ only, 
one can take advantage of the $O(4)$ symmetry and rewrite the ratio of
determinants as a product of multipoles $\nu$
\begin{equation} \label{ScalarDetMult}
  \frac{\det S^{\prime \prime}}{\det {\hat S}^{\prime \prime} } =  
  \prod_{\nu = 1}^\infty \left(\frac{\det S_\nu^{\prime \prime}}{\det 
  {\hat S}_\nu^{\prime \prime} }\right)^{\!d_\nu} \,  ,
\end{equation}
where the fluctuation operator at a fixed $\nu$ is given by
\begin{equation}\label{eq:Spp_def}
  S^{\prime \prime}_\nu = - \partial_\rho^2 - \frac{3}{\rho} \,\partial_\rho + 
  \frac{\nu^2 - 1}{\rho^2} + m^2(\rho) \, ,
\end{equation}
and $d_\nu=\nu^2$ is the `degeneracy factor', {\it i.e.} the number of independent 
multipoles having angular momentum quantum number $\nu$.
The same holds for ${\hat S}^{\prime \prime}$, with $m^2(\rho)$ replaced by $\hat m^2$.
To compute the ratio of determinants, the standard route is to use the 
Gelfand-Yaglom theorem~\cite{Gelfand:1959nq}, which states
\begin{equation} \label{GelYag}
  \frac{\det S^{\prime \prime}_\nu }{\det {\hat S}^{\prime \prime}_\nu } =
  \lim_{\rho \to \infty}\frac{\varphi(\rho)}{\hat{\varphi}(\rho)}\equiv R_\nu \, ,
\end{equation}
where $\varphi$ and $\hat{\varphi}$ functions satisfy respectively 
$S''_\nu\varphi = 0$ and ${\hat S}''_\nu \hat{\varphi} = 0$,
with boundary conditions $\varphi, {\hat\varphi} \sim \rho^{\nu + 1/2}$ 
for $\rho\to 0$. 
This greatly simplifies the problem: instead of computing the full spectrum of the 
$\nu$-fluctuation operator (which is a differential operator in $\rho$), one needs 
to solve the homogeneous differential equations \emph{once}, with specific boundary
conditions.\footnote{For multi-field problems, we need to solve a coupled system of 
differential equations, with $(S''_\nu)_{ij}$ acting on the column vector $\varphi^j$, 
with $j=1,2,...,n$. Imposing regularity conditions at $\rho=0$ (like for $\varphi$ and 
$\hat\varphi$ in~\eqref{GelYag}) we are left with $n$ solutions $\varphi_{(k)}^j$, with 
$k=1,2,...,n$. 
Then $R_\nu=\lim_{\rho\to\infty}\det [\varphi_{(k)}^j(\rho)]/
\det [\hat\varphi_{(k)}^j(\rho)]$, with $\hat\varphi_{(k)}^j$ regular in zero and 
such that $\hat S_{ij}\hat\varphi_{(k)}^j=0$.}
Taking the logarithm of~\eqref{ScalarDetMult} we find
\begin{equation} \label{Rnusum}
  \ln \frac{\det S^{\prime \prime} }{\det {\hat S}^{\prime \prime} } =
  \sum_{\nu=1}^\infty d_\nu \ln R_\nu \, .
\end{equation}

The angular functions that allow for the multipole decomposition of \eqref{scalarSppdef} are 
given in $D=4$ by the Hyper-Spherical Harmonics $Y_{\nu m_A m_B}(\hat x)$.
They are labelled by three indices, with ranges $\nu = 1, 2, \ldots$ and 
$-(\nu-1)/2 \leq m_{A,B} \leq (\nu-1)/2$, from which the degeneracy of 
$d_\nu = (2(\nu-1)/2 + 1)^2 = \nu^2$ follows.
This allows for the expansion of any scalar field configuration according to
\begin{equation}\label{eq:SH_decomp}
  \phi(x) = \sum_{\nu m_A m_B} c_{\nu m_A m_B}(\rho) \ 
  Y_{\nu m_A m_B}(\hat x) \, ,
\end{equation}
with $c_{\nu m_A m_B}$ depending only on the radial variable $\rho=|x|$. 
In this work we do not need explicit expressions for the hyperspherical harmonics, it 
is sufficient to know the group theoretical structure of $SO(4)$.
The decomposition in~\eqref{eq:SH_decomp} corresponds to the statement that, at 
fixed $\rho$, a scalar field is isomorphic, from the point of view of $SO(4)$ rotations, 
to $\bigoplus_{l=0}^\infty (\frac{l}{2},\frac{l}{2})$, with $l=\nu-1$ the orbital 
angular momentum quantum number. 
We adopt the standard notation for $SO(4)$ representations with a pair of half-integer
indices, $(j_A, j_B)$, which characterise the transformation properties under the 
first and second $SU(2)$ factors of $SO(4) \simeq SU(2)_A \otimes SU(2)_B$. 
This is depicted in Fig.~\ref{fig:scalar_fermion}.
The $Y_{\nu m_A m_B}$ are eigenvectors of $J^2\equiv2\left(J^2_A+ J^2_B\right)$, $J_{A_3}$ 
and $J_{B_3}$ with eigenvalues $\nu^2-1$, $m_A$ and $m_B$ respectively. 
We use $J$ to denote the total angular momentum operator and $L$ for the orbital 
angular momentum. 
In the scalar case $J = L$, but for fermions and gauge boson the distinction 
is important, as we will see later on.
Since polarisations do not play an active role apart from accounting for the 
degeneracy factors, from here on we group the indices $m_{A}$, $m_B$ into a single 
index $m$ to avoid clutter.

Because of the spherical symmetry of the problem, the fluctuation operators 
of~\eqref{scalarSppdef} commute with rotations~\cite{Baratella:2024hju}. 
This implies that the action of $S''$ is diagonal in the angular momentum quantum 
numbers, {\it i.e.}
\begin{equation}
  S'' \left[c(\rho)\ Y_{\nu m}(\hat x) \right] = Y_{\nu m}(\hat x) \ 
  S''_\nu\left[ c(\rho) \right]\, ,
\end{equation}
where $S''_\nu$, given in \eqref{eq:Spp_def}, acts only on the radial variable and 
does not depend on polarisations due to spherical symmetry. 
The possibility of reducing the problem to single variable derivative operators 
is what allows the use of the Gelfand-Yaglom theorem, and the multipole 
$\nu$-expansion, which is at the heart of our subtraction procedure.

\vspace{.3cm}
{\bf Dirac notation.} It is convenient to use the $\langle {\bf bra} |$ 
and $| {\bf ket} \rangle$ notation of Dirac to denote different bases for 
the fluctuations, which have the structure of a Hilbert space, and think 
of $S''$ and $\hat S''$ as operators acting on this Hilbert space. 
For example, we consider states $| y_\mu \rangle$ that are eigenvalues of all four 
components of the position operator $X_\mu$, and whose corresponding spatial wave 
function is $\phi(x)\propto \delta^{(4)}(x-y)$.
Of special relevance are bases with definite angular momentum quantum numbers,
denoted by $\ket {\lambda; \nu m}$ and $\ket{\rho; \nu m}$. 
The former $\ket {\lambda; \nu m}$ are eigenvectors of the Laplacian 
operator,\footnote{With a standard abuse of notation, we call $\partial^2$ the abstract 
operator that acts like $\partial^2$ on the (position space) wave functions.}
\begin{align} \label{lameig}
  -\partial^2 \ket {\lambda; \nu m} &= \lambda^2 \ket {\lambda; \nu m} \, , 
  &
  \braket{\lambda; \nu m | \lambda^\prime; \nu^\prime m^\prime} &= 
  \delta(\lambda - \lambda^\prime) \delta_{\nu \nu^\prime} \delta_{m m^\prime} \, ,
\end{align}
while the latter $\ket{\rho; \nu m}$ correspond to a scalar localised on 
the shell $x^2 = \rho^2$ and are normalised as 
\begin{equation}\label{rhoeig}
  \braket{\rho; \nu m | \rho^\prime; \nu^\prime m^\prime} = 
  \rho^{-3} \delta(\rho - \rho^\prime) \delta_{\nu \nu^\prime} \delta_{m m^\prime} \, .
\end{equation}
The scalar product among the elements of the two bases reads
\begin{align}
  \label{angularbasis}
  \braket{\rho; \nu m | \lambda; \nu^\prime m^\prime} &= \frac{\sqrt{\lambda}}{\rho} 
  J_\nu \left( \lambda \rho \right) \, \delta_{\nu\nu^\prime} \delta_{m m^\prime} \, ,
\end{align}
where $J_\nu$ is a Bessel function of the first kind.
These states satisfy the following completeness relations
\begin{equation} \label{complrel}
  1 = \sum_{\nu m} \int_\lambda \ket{\lambda; \nu m} \bra{\lambda;\nu m} 
  = \sum_{\nu m} \int_\rho \ \rho^{3} \ket{\rho; \nu m} \bra{\rho;\nu m} \, .
\end{equation}

\begin{figure}
  \centering
  \begin{tikzpicture}[line width=1.1 pt, scale=1.6, baseline=(current bounding box.center)]
		
	\begin{scope}[shift={(-.2,0)}]	
	\draw[dashed] (-2.5,0) circle  (.5) ;
	\draw[dashed] (-3,0) -- (-4,0) ;
	\draw[black!50!green, fill=white] (-3,0) circle (.08) ;
	\node at (-3.5,-.3) {$ \overline{\phi} $} ;
	\node at (-1.9,.5) {\small $ \phi_i $} ;
	\node[black] at (-3.7,.7) {$T_{1,0}$} ;
	\draw[line width = .5] (-4.1,.4) -- (-3.3,.4) -- (-3.3,.9) ;
	\end{scope} ;
	
	\hspace{.75cm}
	
	\draw[dashed] (0,0) circle  (.5) ;
	\draw[dashed] (-.5,0) -- (-1.5,0) ;
	\draw[black!50!green, fill=white] (-.5,0) circle (.08) ;
	\node[black!40!red] at (.5,0) {\large $ \times $} ;
	\node[black!40!red] at (.75,-.3) { $ \hat{m}^2 $} ;
	\node[black] at (-1.1,.7) {$T_{1,1}$} ;
	\draw[line width = .5] (-4.1+2.6,.4) -- (-3.3+2.6,.4) -- (-3.3+2.6,.9) ;
	
	\hspace{.75cm}

	\begin{scope}[shift={(.3,0)}]
	\draw[dashed] (2.5,0) circle  (.5) ;
	\draw[dashed] (2,0) -- (1,0) ;
	\draw[black!50!green, fill=white] (2,0) circle (.08) ;
	\draw[dashed] (3,0) -- (4,0) ;
	\draw[black!50!green,  fill=white] (3,0) circle (.08) ;
	\node[black!50!green] at (3.5,.3) {$\delta m^2$} ;
	\node[black] at (-1.1+2.5,.7) {$T_{2,0}$} ;
	\draw[line width = .5] (-4.1+5.1,.4) -- (-3.3+5.1,.4) -- (-3.3+5.1,.9) ;
	\end{scope} ;
	
	\draw[white] (5.5,1) -- (5.5,-1);
	\end{tikzpicture}	
  \caption{Diagrammatic expansion of the functional determinant in the number of 
  interaction $\delta m^2$ (green circles) and FV mass $\hat m^2$ (red crosses) insertions. }
  \label{fig:FD_FV}
\end{figure}
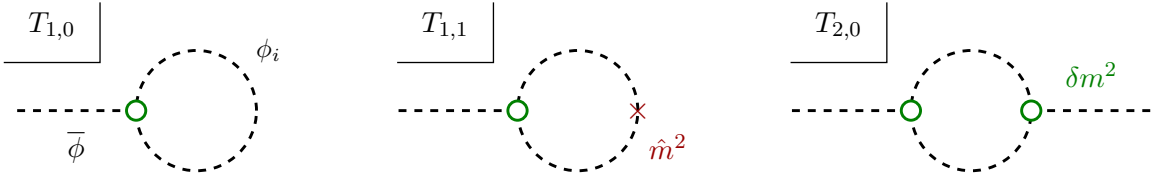
\vspace{.3cm}
{\bf Subtraction procedure.} Our goal is not to provide an explicit solution 
for~\eqref{Rnusum}, but rather to give a simple understanding of how to regularise 
the infinite sum over $\nu$.
We shall characterise the structure of divergences by turning the log of the determinant 
into an infinite sum of traces, which are interpretable in terms of 1-loop Feynman diagrams
with scalar propagators, see Fig.~\ref{fig:FD_FV}.
We begin with the formal operator expansion, where we use~\eqref{scalarSppdef} to write
\begin{equation} \label{lndettrace}
  \ln \frac{\det S^{\prime \prime} }{\det {\hat S}^{\prime \prime} } 
  = \tr \ln \left( 1 + \frac{1}{-\partial^2 + \hat m^2} \delta m^2 \right)=-\sum_{n=1}^\infty 
  \frac{(-1)^{n}}{n}\tr\left( \frac{1}{-\partial^2 + \hat m^2} \delta m^2 \right)^n\, ,
\end{equation}
where $ \delta m^2 \equiv m^2 - \hat m^2$. 
From the diagrammatic point of view, each $(-\partial^2 + \hat m^2)^{-1}$ that comes from 
expanding Eq.~\eqref{lndettrace}  in $n$, corresponds to a propagator,
and each $\delta m^2$ to an insertion of the background field.

A crucial simplification is to organise the entire regularisation procedure in a 
power counting scheme with \emph{two} expansion indices.
Along with the number of insertions of $\delta m^2$, which correspond to the green circles 
in Fig.~\ref{fig:FD_FV}, we expand over the number of mass insertions $\hat m^2$, denoted 
by the red crosses in Fig.~\ref{fig:FD_FV}.
In performing this double expansion we get
\begin{equation} \label{Tracexpand}
\begin{split}
  \ln \frac{\det S^{\prime \prime} }{\det {\hat S}^{\prime \prime} }   
  &= \tr \frac{1}{-\partial^2} \delta m^2{-}\tr \frac{1}{-\partial^2} \hat m^2  
  \frac{1}{-\partial^2} \delta m^2{-}\frac{1}{2} \tr  \frac{1}{-\partial^2} \delta m^2 
  \frac{1}{-\partial^2} \delta m^2 + O\big((\delta m^2)^3,\hat{m}^4\big)
  \\
  &\equiv T_{1,0} - T_{1,1} - \frac{1}{2} T_{2,0} + O\big((\delta m^2)^3,\hat{m}^4\big) \, .
 \end{split}
\end{equation}
We have labelled the individual traces by $T_{i, j}$, where the first index $i$ counts the
number of interactions $\delta m^2$, and the second index $j$ the number of mass insertions
$\hat m^2$, as in Fig.~\ref{fig:FD_FV}.
The relation to Feynman diagrams makes it easy to determine the degree of divergences in
any $D$ spacetime dimensions.
In $D = 4$, we have a quadratic divergence carried by $T_{1,0}$ and a logarithmic divergence 
in $T_{1,1}$ and $T_{2,0}$. 
In $D = 3$, there is only a linear divergence in $T_{1,0}$.
This is obvious if one thinks about the one-loop integral in momentum space and counts 
the number of scalar propagators in the loop. 
We will see that in the angular momentum basis the same UV divergence structure is evident 
at large multipoles $\nu$.

We are going to evaluate these divergent traces via insertions of the appropriate states in 
the angular momentum bases. 
This will provide a series that matches the large $\nu$ behaviour of~\eqref{Rnusum}.
Subtracting it from from $\sum_{\nu=1}^\infty d_\nu \ln R_\nu$ will make the series
convergent.
Then we will evaluate the same traces in momentum space using dimensional regularisation
and add them back to obtain our final expression. 
In short, we subtract the divergent pieces and add back their regularised counterparts
\begin{equation} \label{eq:DetS}
  \ln \frac{\det S^{\prime \prime} }{\det {\hat S}^{\prime \prime} } = 
  \sum_{\nu = 1}^\infty d_\nu \ln R_\nu - 
  \left( T_{1,0} - T_{1,1} - \frac{1}{2} T_{2,0} \right) + 
  \left( T^{\rm DR}_{1,0} - T^{\rm DR}_{1,1} - \frac{1}{2} T^{\rm DR}_{2,0} \right) \, ,
\end{equation}
where DR stands for dimensional regularisation.

Let us start with the evaluation in angular momentum space.
$T_{1,0}$ consists of a single insertion of the interaction, a massless propagator and 
no mass insertions:
\begin{equation} \label{T10s}
\begin{split}
  T_{1,0} &= \sum_{\nu m} \int_\rho \ \rho^3 \sum_{\nu^\prime m^\prime} 
  \int_\lambda \bra{\rho; \nu m} \frac{1}{-\partial^2} \ket{\lambda; \nu^\prime m^\prime} 
  \bra{\lambda; \nu^\prime m^\prime} \delta m^2 \ket{\rho; \nu m} 
  \\
  &= \sum_{\nu m} \int_\rho \ \rho^3 \int_\lambda \frac{1}{\lambda^2}  
  \frac{\sqrt{\lambda}}{\rho} J_\nu(\lambda \rho) \bra{\lambda; \nu m} \delta m^2 \ket{\rho; \nu m} 
  \\
  &= \sum_{\nu m} \int_\rho \ \rho^3 \delta m^2 \int_\lambda \frac{1}{\lambda^2} 
  \left( \frac{\sqrt{\lambda}}{\rho} J_\nu(\lambda \rho)  \right)^2 
  = \sum_\nu d_\nu \frac{1}{2\nu} \int_\rho \ \rho \ \delta m^2 \, .
\end{split}
\end{equation}
In the first line we used the completeness relations from~\eqref{complrel}, in the 
second we used~\eqref{lameig} and~\eqref{angularbasis} and in the third 
again~\eqref{angularbasis}, together with the fact that the operator $\delta m^2$
acts as multiplication by $\delta m^2(\rho)$ in the $|\rho;\nu m\rangle$ basis\footnote{
To avoid clutter we shall suppress the explicit $\rho$ dependence in the functions
and use the same descriptor $\delta m^2$ for the operator and the function and reinstate
explicit arguments only when needed.}.
In the last step we performed the integral over $\lambda$ using
$\int_0^\infty \frac{{\rm d} x}{x}  J^2_\nu(x) = \frac{1}{2\nu}$,
and finally summed over the compound index $m$ to obtain the degeneracy factor $d_\nu$.
The final expression for $T_{1,0}$ is pretty simple; the integral over $\rho$ is finite, 
as $\delta m^2 \xrightarrow{\rho \to \infty} 0$ faster than $\rho^{-2}$ in all cases of 
physical interest.
In $D = 4$, we have $d_\nu = \nu^2$ and the sum $\sum_\nu \nu$ diverges quadratically 
when $\nu \to \infty$.
This is the anticipated UV quadratic divergence in the angular momentum decomposition 
of $T_{1,0}$. 

Next, we compute $T_{1,1}$, which has an additional mass insertion
\begin{equation}\label{eq:T11_scal}
\begin{split}
  T_{1,1} &= \sum_{\nu m} \int_\rho \ \rho^{3}  
  \int_\lambda \bra{\rho; \nu m} \frac{\hat m^2}{\left( -\partial^2 + \delir^2 \right)^2}  
  \ket{\lambda; \nu m} \bra{\lambda; \nu m} \delta m^2 \ket{\rho; \nu m} 
  \\
  &= \sum_{\nu} d_\nu \ \hat m^2 \int_\rho \ \rho \  \delta m^2
  \int_\lambda \frac{\lambda}{\left( \lambda^2 + \delir^2 \right)^2} J_\nu^2(\lambda \rho) \, .
\end{split}
\end{equation}
We dropped the sum over $\nu'm'$ and set all labels equal, because all operators
are rotationally symmetric so they do not mix states with different $\nu m$.
Since $J_\nu(x \to 0) \sim x^\nu$, the integral over $\lambda$ has an IR divergence 
for $\nu \leq 1$. 
We regulate it with the dimensionful parameter $\delir$, which will cancel out later.
In $D = 4$ the lowest value of $\nu$ is 1, which involves the IR-divergent integral
\begin{equation}\label{eq:T11_IR}
  \int_\lambda \frac{\lambda}{ \left( \lambda^2 +\delir^2 \right)^2} J_1^2(\lambda \rho) 
  = -\frac{\rho^2}{16} \left( 1 + 4 \gamma_E + 4 \ln \frac{\rho \delir}{2} \right) + 
  {\cal O}(\delir^2) \, .
\end{equation}
For $\nu > 1$ the integration over $\lambda$ produces no IR divergence, so we set 
$\delir = 0$ and get
$\int_\lambda 1/\lambda^3 J_\nu^2(\lambda\rho) = \rho^2/(4\nu \left( \nu^2 - 1 \right))$.
Combining both pieces we have
\begin{equation} \label{T11angbad}
  T_{1,1}  = \frac{1}{4} \int_\rho \ \rho^3 \delta m^2 \hat m^2 \left( \sum_{\nu > 1} 
  d_\nu \frac{1}{ \nu \left( \nu^2 - 1 \right)} - \frac{1}{4} - {\gamma_E} - 
  \ln \frac{\rho\delir}{2} \right) \, .
\end{equation}
In the first term we see that the summand goes as $1/\nu$ at large $\nu$, which is the 
anticipated UV logarithmic divergence. 
We could keep $T_{1,1}$ in this form, however, it is useful to unify all the subtraction
terms to a common form that removes the minimal amount of high-$\nu$ divergence to 
make the sum in~\eqref{eq:DetS} finite.
We thus add and subtract $\sum_{\nu=2}^\infty d_\nu/\nu^3$ and account for the 
finite difference with 
$\sum_{\nu = 2}^\infty \nu^2 \left( \frac{1}{\nu(\nu^2-1)} - \frac{1}{\nu^3} \right) = 1/4$,
such that
\begin{equation} \label{T11ang}
\begin{split}
  T_{1,1} &= \frac{1}{4}\int_\rho \ \rho^3 \delta m^2 \hat m^2 \left( \sum_{\nu=1}^\infty 
  d_\nu \frac{1}{\nu^3} -1 - {\gamma_E} - \ln \frac{\rho\delir}{2} \right)  \, ,
  \end{split}
\end{equation}
where the sum starts at $\nu = 1$, which is compensated by adding $-1$.
This form isolates the log divergence in the simplest term 
$\sum_{\nu=1}^\infty d_\nu/\nu^3$ and has the advantage that the sum starts at $\nu = 1$, 
as in the other $T_{i,j}$'s.
Moreover, the remaining finite pieces will combine nicely with those from the DR parts, as 
we will see soon.

The final divergent trace $T_{2,0}$ comes with two interactions and no mass insertion
\begin{equation} \label{T20fac2}
\begin{split}
  T_{2, 0} &= \sum_{\nu m}
  \int_\rho \rho^{3} \int_{\lambda} \int_{\rho'} \rho^{\prime 3} 
  \int_{\lambda'} \bra{\rho;\nu m} \frac{1}{-\partial^2} \ket{\lambda; \nu m} 
  \\
  & \qquad \bra{\lambda; \nu m} \delta m^2 
  \ket{\rho'; \nu m} \bra{\rho';\nu m} \frac{1}{-\partial^2}   
  \ket{\lambda'; \nu m} \bra{\lambda';\nu m} \delta m^2 \ket{\rho; \nu m} 
  \\
  &= \sum_{\nu m} \int_\rho \ \rho \ \delta m^2 \int_{\rho'} \ 
  \rho' \ \delta m^2(\rho')
  \int_{\lambda} \frac{J_\nu(\lambda \rho) J_\nu(\lambda_1 \rho')}{\lambda}
  \int_{\lambda'}  \frac{J_\nu(\lambda' \rho) J_\nu(\lambda' \rho')}{\lambda'}  
  \\
  &= 2 \sum_{\nu m} \frac{1}{4\nu^2} \int_\rho \ \rho \ \delta m^2 
  \int_0^\rho {\rm d} \rho' \ \rho' \ \delta m^2(\rho')  \left( 
  \frac{\rho'}{\rho} \right)^{2 \nu}  \, ,
 \end{split}
\end{equation}
where we used
\begin{align}\label{eq:rho1rho2}
  \int_0^\infty {\rm d} \lambda  \frac{1}{\lambda} J_\nu(\lambda \rho) J_\nu(\lambda \rho') &= 
  \frac{1}{2\nu} \left( \frac{\rho'}{\rho} \right)^\nu \, , & {\rm for} \ \rho' &< \rho \, .
\end{align}
The factor of 2 in the last equality of~\eqref{T20fac2} comes about because the two regions $\rho' < \rho$ and 
$\rho < \rho'$ contribute the same amount.
Defining $x \equiv \rho' / \rho$ and summing over $m$, we have
\begin{equation}\label{T20ang}
  T_{2, 0} = \sum_\nu d_\nu \frac{1}{2\nu^2} \int_\rho \ \rho^3 \delta m^2
  \int_0^1 {\rm d} x \ x^{2 \nu + 1} \delta m^2(x \rho) \, .
\end{equation}
To examine the UV behaviour, note that the integral over $x$ at large $\nu$ is
\begin{equation}\label{xnuintegral}
  \int_0^1 {\rm d}x \ x^{2\nu+1}  \delta m^2(x \rho) = \frac{1}{2\nu} \delta m^2 
  + O(\nu^{-2}) \, ,
\end{equation}
where we Taylor-expanded $\delta m^2(x \rho) = \delta m^2(\rho)+(x-1)
\frac{\rm d}{{\rm d} x}\delta m^2(\rho x)|_{x=1}+\ldots$. 
This shows that the summand in $T_{2,0}$ goes as $d_\nu/\nu^3=1/\nu$, whose series 
is logarithmically divergent.
The observation in~\eqref{xnuintegral} suggests the following separation $\delta m^2(x\rho)=
\delta m^2(\rho)+(\delta m^2(x\rho)-\delta m^2(\rho))$, such that $T_{2,0}$ is decomposed into (i) a term proportional to 
$\left(\delta m^2(\rho)\right)^2$, where the $\int {\rm d}x$ integral can be trivially
performed, yet is divergent as a series in $\nu$, and (ii) a term that is convergent 
in $\nu$ so the series can be summed explicitly. 
Performing the ${\rm d}x$ integration on one side and the $\sum_\nu$ on the 
other,~\eqref{T20ang} becomes
\begin{equation}
 T_{2,0} = \sum_\nu \frac{d_\nu}{4\nu^2(\nu + 1)} \int_\rho \rho^3 
 \left( \delta m^2 \right)^2 + \frac{1}{2} \int_\rho \rho^3 \delta m^2
 \int_0^1 {\rm d}x \frac{x^3}{1 - x^2} \left( \delta m^2(x \rho) - \delta m^2 \right) \, .
\end{equation}
The first term encodes the logarithmic UV divergence, but still contains a finite part. 
It is useful, as we did for $T_{1,1}$ above, to isolate the divergence by adding and 
subtracting $1/(4\nu)$ in the sum, and use
$\sum_{\nu=1}^\infty ( \frac{1}{\nu+1} - \frac{1}{\nu}) = -1$ to obtain
\begin{equation}\label{eq:T20final}
\begin{split}
 T_{2,0} 
 & = \sum_\nu \frac{d_\nu}{4\nu^3} \int_\rho \rho^3 
 \left( \delta m^2 \right)^2 -\frac{1}{4} \int_\rho \rho^3 \left( \delta m^2 \right)^2 
 \\
 & \qquad 
  + \frac{1}{2} \int_\rho \rho^3 \delta m^2
 \int_0^1 {\rm d}x \frac{x^3}{1{-}x^2} \left( \delta m^2(x \rho) - \delta m^2\right) \, .
\end{split}
\end{equation}
Apart from the first term, which is divergent, the remaining pieces are finite. 
They will combine and simplify against analogous terms coming from $T_{2,0}^{\rm DR}$.

We found simple expressions for $T_{1,0}$, $T_{1,1}$, and $T_{2,0}$ by using 
angular momentum bases with definite $\lambda$ and $\rho$. 
The manipulations we performed can be extended to generic $D$ dimensions, and with 
this method one can also compute higher order terms in the expansion~\eqref{Tracexpand}.
In $D = 4$, the terms we obtained above are sufficient as they encode 
all the UV divergences, while for higher $D$ one would have to subtract more terms.
We are done with computing the divergent pieces in angular momentum space and we can move
on to dimensionally regularised traces in momentum space. 

We start with $T^{\rm DR}_{1,0}$ and resolve the trace defining $T_{1,0}$ using the usual 
position-$x$ and momentum-$q$ bases:
\begin{equation} \label{T11scaless}
\begin{split}
  T^{\rm DR}_{1,0} & = \tr \frac{1}{-\partial^2} \delta m^2 
  = \int_x \int_q \bra{x} \frac{1}{-\partial^2} \ket{q} \bra{q} \delta m^2 \ket{x} 
  \\
  &= \int_x \int_q \frac{e^{i q\cdot x}}{q^2} \delta m^2(|x|) e^{-i q\cdot x} 
  = \widetilde{\delta m^2}(0) \int_q \frac{1}{q^2} \, ,
\end{split}
\end{equation}
where a tilde denotes the Fourier transform, defined as 
$\tilde{f}(q) = \int_x \, e^{iq\cdot x} f(x)$.
The integral in \eqref{T11scaless} is scaleless and vanishes in DR, {\it i.e.}
$T^{\rm DR}_{1, 0} = 0$.

The trace with two insertions $T^{\rm DR}_{1,1}$ is evaluated using DR with 
$D = 4 - \epsilon$ and keeping the regulator $\delir$
\begin{equation} \label{eq:T11Reg}
\begin{split}
  T^{\rm DR}_{1,1} & = \hat m^2 \ \widetilde{\delta m^2}(0) \ \mu^\epsilon 
  \int \frac{{\rm d}^D q}{\left( 2 \pi \right)^D} 
  \frac{1}{\left( q^2 + \delta_{\rm IR}^2 \right)^2} = 
   \hat m^2 \ \widetilde{\delta m^2}(0) 
   \frac{\Gamma \left( \frac{\epsilon}{2} \right)}{(4\pi)^{2-\frac{\epsilon}{2}}} 
   \left( \frac{\mu}{\delta_{\rm IR}} \right)^\epsilon \, .
 \end{split}
\end{equation}
Rewriting $\widetilde{\delta m^2}(0)$ as an integral in position space (doing the Fourier
transform at $q = 0$), and in the limit of $\epsilon,\delir\to 0$, we find
\begin{equation}\label{eq:T11Reg2}
   T^{\rm DR}_{1,1}= \frac{\hat m^2 }{8} \int_\rho \ \rho^3 \delta m^2 \left( 
   \frac{2}{\epsilon} + 2 \ln \frac{\tilde \mu}{\delta_{\rm IR}} \right) \, .
\end{equation}
The IR regulator $\delta_{\rm IR}$ cancels against the one in~\eqref{T11ang} and the final 
result does not contain any IR divergence.
The trace was evaluated in momentum space and we ended up with a simple expression 
that isolates the $1/\epsilon$ and $\ln \mu$ dependence and a finite part, with an 
overall coefficient that is obtained by a simple integration of $\delta m^2$ over $\rho$.

Let us turn to $T_{2,0}^{\rm DR}$, where we have
\begin{equation} \label{T20regp1}
\begin{split}
  T^{\rm DR}_{2,0} &= \int_x\int_p \int_y\int_q
  \bra{x} \frac{1}{-\partial^2} \ket{q} \bra{q} \delta m^2 \ket{y} \bra{y} 
  \frac{1}{-\partial^2} \ket{p} \bra{p} \delta m^2 \ket{x} 
  \\
  &= \int_p \frac{1}{p^2} \int_q \frac{1}{q^2} 
  \int_y e^{i y \cdot(p-q)} \delta m^2(y) \int_x e^{i x \cdot (q-p)} \delta m^2(x) 
  \\
  &= \int_k  \widetilde{\delta m^2}(k) \widetilde{\delta m^2}(-k) \mu^\epsilon 
  \int \frac{\text{d}^D q}{\left( 2 \pi \right)^D} \frac{1}{q^2(k+q)^2} 
  \\
  &= \frac{1}{16\pi^2} \int_k \widetilde{\delta m^2}(k) \widetilde{\delta m^2}(-k) 
  \left( \frac{2}{\epsilon} 
  + 2 + \ln \frac{\tilde \mu^2}{k^2} \right) \, .
\end{split}
\end{equation}
From the second to the third line we introduced $k = p - q$, switched the integral over 
$q$ from 4 to $D = 4 - \epsilon$ dimensions and integrated using dimensional regularisation.
The final line contains, in round brackets, two pieces with different $k$ dependences.
The first is a constant and is converted back to an integral over $\rho$ in position space, just as in 
$T^{\rm DR}_{1,1}$ above.
The second term requires more work. 
In Appendix~\ref{app:FTProofs} we derive the following identity, valid for any regular function 
$f(\rho)$ depending only on the Euclidean radius $\rho$:
\begin{equation}\label{eq:Fourier}
\begin{split}
  \int_k \widetilde{f}( \left| k \right|)^2 \ln k^2 & = 
  -4 \pi^2 \int_\rho \rho^3 f(\rho)\left( f(\rho) \left( \gamma_E{-}1{+}\ln \frac{\rho}{2} \right) + 
  2 \int_0^1 {\rm d} x\, x^3\, \frac{f(x \rho) - f(\rho)}{1-x^2} \,\right) ,
\end{split}
\end{equation}
With this we can rewrite~\eqref{T20regp1} as
\begin{equation} \label{eq:T20Reg}
\begin{split}
  T^{\rm DR}_{2,0} &= \frac{1}{8} \int_\rho \ \rho^3 \left( \delta m^2 \right)^2 
  \left( \frac{2}{\epsilon} 
  + 2 \gamma_E + 2 \ln \frac{\tilde \mu \rho}{2} \right)  
  \\
  & \quad +\frac{1}{2} \int_\rho \ \rho^3 \delta m^2 \int_0^1 {\rm d} x 
  \frac{x^3}{1 - x^2} \left( \delta m^2(x \rho) - \delta m^2 \right) \, .
\end{split}
\end{equation}
The second line matches exactly the last term in~\eqref{eq:T20final}, so the two cancel out.
Combining everything in~\eqref{eq:DetS} we find a very compact expression
\begin{equation} \label{eq:DetS_MMSFin}
 \begin{split}
 \ln \frac{\det S^{\prime \prime} }{\det {\hat S}^{\prime \prime} } 
  &= \sum_{\nu = 1}^\infty \left( d_\nu \ln R_\nu - \frac{\nu}{2} 
  \int_\rho \, \rho \, \delta m^2 
 + \frac{1}{8 \nu} \int_\rho \, \rho^3 \,\delta m^4  \right)
 \\
 &- \frac{1}{8}  \int_\rho  \,\rho^3 \,\delta m^4  
  \left( \frac{1}{\epsilon} 
  + 1 + \gamma_E + \ln \frac{\tilde \mu \rho}{2}  \right) \, ,
\end{split}
\end{equation}
which matches~\eqref{eq:DetS_scal}.
The terms with $\hat m^2\delta m^2$ and $(\delta m^2)^2$ nicely combine into
$\delta m^2(\delta m^2 + 2 \hat m^2) = m^4 - \hat m^4 \equiv \delta m^4$.

Let us finally comment on the inclusion of more scalars and how to go 
from~\eqref{eq:DetS_MMSFin} to~\eqref{eq:DetS_scal}.
One has to include states $|\lambda;\nu m;i\rangle$ with the flavour index $i$ (and 
similarly for $|\rho \rangle$).
The operator $\delta m^2$ now acts like 
$\delta m^2|\rho;\nu m;i\rangle = |\rho;\nu m;j\rangle\delta m^2_{ji}(\rho)$, mixing the 
various flavours.
The rest is essentially the same, with a leftover `$\tr$' over the flavour indices 
instead of the whole Hilbert space. 
Notice that there is no need to diagonalise the matrix $\delta m^2_{ij}(\rho)$, nor 
$\hat m^2_{ij}$, because they always appear as interaction vertices in the 
numerator.\footnote{On the contrary, what appears in the denominator of~\eqref{Tracexpand}, 
{\it i.e.} $\partial^2$, needs to be diagonalised in order to deal with its inverse. 
The addition of flavour does not spoil the diagonal structure of $\partial^2$ in 
$\lambda$ space, as $\partial^2|\lambda;\nu m;i\rangle=-\lambda^2|\lambda;\nu m;i\rangle$.}

\vspace{.3cm}
{\bf Previous work.}
It is interesting to compare our result for scalar fluctuations to previous literature. 
In Ref.~\cite{Baacke:2003uw}, the authors considered the same problem and also used Feynman 
diagrams and dimensional regularisation. 
Compared to their work we perform an additional expansion in $\hat m^2$, see~\eqref{Tracexpand}.
This is key to arrive at our simpler recipe with single integrals, while in 
Ref.~\cite{Baacke:2003uw} double integrals remain.
In Ref.~\cite{Dunne:2006ct}, the authors use the $\zeta$-function regularisation method and arrive 
at the same result.
Comparing to them\footnote{In the notation of Ref.~\cite{Dunne:2006ct}, the $V$ corresponds to 
our $\delta m^2$ while $V(V+ 2m^2)$ to $\delta m^4$.}
we believe that our method has some merits: (i) Feynman diagrams and dimensional regularisation 
are more familiar to the wide particle physics community, (ii) we generalise it to fluctuations 
with fermions and gauge bosons and (iii) it can be connected more easily to the other part of 
the calculation of false vacuum decay, where one has to compute counterterms and running of 
the couplings in the bounce action.
This is also typically done with Feynman diagrams and dimensional regularisation.   

%
%
\section{Fermions} \label{sec:fermion}
In this section we focus on fermions and work out the UV subtraction and regularisation
procedure.
We start with the simplest Yukawa action in Euclidean signature for a real scalar and a 
Dirac fermion
\begin{equation}
  S[\phi,\psi] = S[\phi] + 
  \int_x \left(\bar{\psi}i\gamma_\mu\partial_\mu \psi + 
  \left(m + y \phi \right) \bar{\psi} \psi\right) \, .
\end{equation}
At the end we consider the general case with an arbitrary number of flavours and 
include the pseudo-scalar terms of the type $\phi\bar{\psi}\gamma_5\psi$, as well
as Majorana fermions. 
The scalar action $S[\phi]$ is like in Section~\ref{sec:scalar} and features a metastable 
FV $\hat\phi$ and a bounce solution $\bar\phi(\rho)$.

We want to compute the ratio of determinants among the fluctuation operators on the bounce
and the FV.
Following the notation of Section~\ref{sec:summary}, we have 
$S'' = \cancel{\partial} + m_\psi(\rho)$, while $\hat S'' = \cancel{\partial}+\hat m_\psi$, 
where $m_\psi = m + y \bar\phi$ and $\hat m_\psi = m + y \hat\phi$. 
Given that the determinants of $S''$ and its hermitian conjugate $S''^\dagger$ are the 
same~\cite{Zinn-Justin:572813} (and similarly for $\hat S''$), we have~\cite{Isidori:2001bm}
\begin{equation}\label{fermdet}
  \frac{\det S''}{\det \hat S''} =
  \sqrt{\frac{\det S''^\dag S''}{\det \hat{S}''^\dag \hat{S}''}} = 
  \sqrt{\frac{\det \left(-\partial^2+ m_\psi^2(\rho)-\cancel{\partial}m_\psi \right)}{
  \det \left(-\partial^2+ \hat{m}_\psi^2\right)}} \, .
\end{equation}
To work out the product among the differential operators we have used that 
$(\cancel{\partial})^\dagger=-\cancel{\partial}$~\cite{Zinn-Justin:572813}.

{\bf Multipole decomposition.} Like for the scalar field, our study of the 
fluctuation operators starts from a choice of basis on the function space 
$\psi_\alpha(x)$ of spinor configurations, that is the space where the operators act on. 
Given the rotationally symmetric nature of the problem, the fluctuation operators on both 
the FV and the bounce commute with the total angular momentum generators, 
{\it i.e.} $[ J_{\mu\nu}, S'' ] = 0$. 
Therefore, we look for a basis of `Dirac-spinor spherical harmonics' 
${Y}_\alpha^{(\nu M)}(\hat x)$ that are $SO(4)$ multiplets, and with the property that 
any Dirac spinor $\psi_\alpha(x)$ can be decomposed as 
$\psi_\alpha(x)=\sum_{\nu, M}c_{\nu M}(\rho){Y}_\alpha^{(\nu M)}(\hat x)$. 
Here, $\nu$ is a quantum number that characterises total angular momentum, while $M$ 
collects all other quantum numbers (polarisations) that are needed to fix the state 
at given $\nu$.
 
\begin{figure}[t]
  \centering
  \begin{tikzpicture}[line width=1.1 pt, scale=.7, baseline=(current bounding box.center)]
		\draw[-stealth] (0,0) to  (0,5.5) ;
		\draw[-stealth] (0,0) to  (5.5,0) ;
		
		\node at (0,6) {$j_B$};
		\node at (6,0) {$j_A$};
				
		\foreach \x in {0,1,2}
		{\node at (2*\x,-.5) {\footnotesize $\x$} ;
		\node at (-.7,2*\x) {\footnotesize $\x$} ;}
		\foreach \x in {1,3,5}
		{\node at (\x,-.5) {\footnotesize $\frac{\x}{2}$} ;
		\node at (-.7,\x) {\footnotesize $\frac{\x}{2}$} ;}
		
		\filldraw[black] (0,0) circle (.5pt) ;
		\filldraw[black] (0,1) circle (.5pt) ;
		\filldraw[black] (1,0) circle (.5pt) ;

  		\foreach \x in {1,2,3,4,5} \foreach \y in {1,2,3,4,5}
  		\filldraw[black] (\x,\y) circle (.5pt) ;
		
		\foreach \x in {0,1,2,3,4,5}
    		\draw[black!40!red,line width =.8] (\x,\x) circle (5pt) ;
		
		\foreach \x in {0,1,2,3,4}
    		{\draw[black!20!blue,line width =.6] (\x,\x+1) circle (4pt) ;
		\draw[black!20!blue,line width =.6] (\x,\x+1) circle (7pt) ;
		\draw[black!20!blue,line width =.6] (\x+1,\x) circle (4pt) ;
		\draw[black!20!blue,line width =.6] (\x+1,\x) circle (7pt) ;}
		
		\node[black!40!blue] at (3.12,1.5) {\tiny $\chi=\pm 1$} ;

		\node[black!40!red] at (8,3.5) {\small Scalar $\phi(\hat x)$ };
		\node[black!20!blue] at (8.07,2.7) {\small Dirac $\psi_\alpha(\hat x)$};		
\end{tikzpicture}
  \vspace{.2cm}
  \caption{$SO(4)$ irreducible representations lattice. 
  Red circles represent the $SO(4)$ multiplets into which a scalar function $\phi(\hat x)$ 
  decomposes, corresponding to a multiplet of (hyper)spherical harmonics with fixed total
  angular momentum.
  Blue circles are the multiplets making up a Dirac spinor $\psi_\alpha(\hat x)$; double
  circles cover the two chiralities $\chi = \pm 1$. }
  \label{fig:scalar_fermion}
\end{figure}
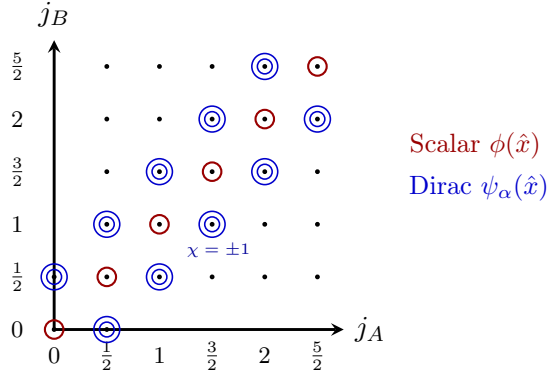

To characterise the ${Y}_\alpha^{(\nu M)}(\hat x)$ basis, we consider the Dirac spinor
$\psi_\alpha(\hat x)$ (we suppress the radial variable as it is inert under rotations)
from the point of view of rotations, as a tensor product among a scalar function in 
the $\hat x$ variable and a $(1/2, 0)\oplus(0, 1/2)$ finite-dimensional 
representation of $SO(4)$. 
A scalar function is an infinite dimensional representation 
$\bigoplus_{l=0}^\infty \left(l/2, l/2 \right)$, as in the previous section.
Working out the tensor product among $(1/2,0)$ and $(l/2, l/2)$ with $l > 0$ using the 
tensor product rules familiar from $SO(3)$, we find
\begin{equation}
  \left( \tfrac{1}{2},0 \right) \otimes \left( \tfrac{l}{2}, \tfrac{l}{2} \right)
  \simeq \left( \tfrac{l+1}{2}, \tfrac{l}{2} \right) \oplus 
  \left( \tfrac{l-1}{2},\tfrac{l}{2} \right) \, ,
\end{equation}
and a symmetric result for $(0, 1/2)$. 
For $l = 0$, the tensor product is trivial and we end up with only one term. 
After summing over $l$, we can rearrange the infinite sum as
\begin{equation}\label{eq:group_th_spinor}
  \left( \left( \tfrac{1}{2}, 0 \right) \oplus \left( 0, \tfrac{1}{2} \right) \right)
  \, \otimes \, \bigoplus_{l = 0}^\infty\ \left( \tfrac{l}{2}, \tfrac{l}{2} \right)\ 
  \simeq \ 2 \, \bigoplus_{j=0}^\infty \left( \left(\tfrac{j}{2},\tfrac{j+1}{2} \right)
  \oplus \left( \tfrac{j+1}{2}, \tfrac{j}{2} \right) \right) \, ,
\end{equation}
where the factor of 2 in front of everything on the RHS means that there are two copies 
for each representation. 
This is depicted in Fig.~\ref{fig:scalar_fermion}, where the notation $(j_A,j_B)$ is 
used for the `coordinates' of an irreducible representation in the 
$SO(4) \simeq SU(2)_A \otimes SU(2)_B$ representation lattice. 
A way to distinguish the two equivalent representations is to notice that each copy 
comes from either the left-chiral component of $\psi_\alpha$, {\it i.e.} the one 
transforming as $(1/2,0)$, or from the right-chiral one $(0,1/2)$. 
An operator that distinguishes the two is $\gamma_5$, so we can label the two copies 
according to the $\gamma_5$ eigenvalue $\chi = \pm 1$. 
Notice however that $\gamma_5$ does not commute with $S''^\dagger S''$ due to the term
$\cancel\partial m_\psi$, so we expect this last term to induce mixing among states with 
different chirality $\chi$ and the same angular momentum quantum numbers.

This discussion motivates the introduction of a basis $Y_\alpha^{(\nu\sigma m \chi)}(\hat x)$, 
such that any spinor can be decomposed as
\begin{equation}
  \psi_\alpha(\rho,\hat x) = \sum_{\nu\sigma m \chi}
  c_{\nu\sigma m \chi}(\rho) \, Y_\alpha^{(\nu\sigma m \chi)}(\hat x) \, .
\end{equation}
We choose $\nu \equiv j_A {+} j_B+1/2$ to label the total angular momentum, 
$\sigma = 2( j_A {-} j_B)=\pm1$ selects the lower/upper diagonal in Fig.~\ref{fig:scalar_fermion}, $m$ collects all the polarisations, of which there 
are $d_\nu^\psi = \nu(\nu+1)$, and $\chi$ labels the two possible chiralities.
The virtue of this basis is that a rotational invariant operator, such as 
$S''^\dagger S''$, can only mix states with the same $\nu, \sigma$ and $m$. 
Moreover, $SO(4)$ symmetry implies that the effect of such an operator does not 
depend on $m$. 
Invariance under the exchange of $j_A$ and $j_B$ also implies that 
we can set $\sigma = +1$ and multiply the end result by 2.

{\bf Action of the fluctuation operator.}
Previous arguments imply that the action of the operator $S''^\dagger S''$ on a state 
with definite angular momentum quantum numbers is of the following form
\begin{equation}
 S''^\dagger S'' \left[ c(\rho)Y_\alpha^{(\nu\sigma m\chi)}(\hat x) \right] =
 \sum_{\chi'} Y_\alpha^{(\nu\sigma m\chi')}(\hat x) \, 
 \left( S''^\dagger S'' \right)^{(\nu)}_{\chi\chi'} \left[c(\rho)\right] \, ,
\end{equation}
where the partial operator $\left(S''^\dagger S''\right)^{(\nu)}$ only depends on $\nu$
and not on $m$, due to rotational symmetry.
In particular, an explicit computation similar to~\cite{AVAN1986621,Isidori:2001bm} gives
\begin{equation}\label{blockDdagD} 
 \left(S''^\dag S''\right)^{(\nu)} = 
 \begin{pmatrix} 
 -\Delta_{\nu-1}+m_\psi^2   & -\dot m_\psi 
 \\ 
 -\dot m_\psi               & -\Delta_{\nu} + m_\psi^2
 \end{pmatrix} \, ,
\end{equation}
where we introduced the differential operator
\begin{equation} \label{eq:def_Dj}
  \Delta_\nu = \partial_\rho^2 + \frac{3}{\rho} \partial_\rho - \frac{\nu(\nu+2)}{\rho^2} \, .
\end{equation}
The form of this $\nu$-partial differential operator in $\rho$ that couples the two 
chiralities, can be understood purely in group theoretical terms. 
By inspecting the diagonal terms, we see that the chiral states also have definite 
orbital angular momentum $L^2 = L_{\mu\nu} L_{\mu\nu}$, with eigenvalues $\nu(\nu+2)$ 
and $(\nu-1)(\nu+1)$. 
This could have been anticipated from~\eqref{eq:group_th_spinor}: each pair of equivalent
representations $((j+1)/2, j/2)$ comes from orbital momentum eigenstates with
either $l = j + 1$ or $l = j$, whose $L^2$ eigenvalues are precisely $\nu(\nu + 2)$ 
and $(\nu - 1)(\nu + 1)$. 
The off-diagonal terms can be understood after noting that $\cancel{\partial}m_\psi$ 
can be expressed as $\gamma_\mu \hat x_\mu\dot m_\psi$, and that 
$\{\gamma_5,\gamma_\mu \hat x_\mu\} = 0$ and 
$(\gamma_\mu \hat x_\mu)^2 = 1$, meaning that $\gamma_\mu \hat x_\mu$ flips the 
eigenvalue of $\gamma_5$ when applied once and returns the initial state when applied twice.

We have focused so far on the angular properties of states. 
To completely fix a basis for Dirac-spinor fluctuations we consider elements with 
a definite radial location $|\rho; a\rangle$, where $a$ collects all discrete quantum numbers $\nu,m,\sigma,\chi$, and elements with definite 
$\partial^2 = -\lambda^2$, {\it i.e.} $|\lambda; a\rangle$, as we did for scalars above.
These are normalised respectively as 
$\langle \rho; a|\rho'; a'\rangle=\rho^{-3}\delta(\rho-\rho')\delta_{aa'}$
and 
$\langle \lambda; a|\lambda'; a'\rangle=\delta(\lambda-\lambda')\delta_{aa'}$. 
They satisfy
\begin{equation}\label{eq:rholambda}
  \left \langle \rho; \nu \sigma m \chi | \lambda;\nu' \sigma' m' \chi' \right \rangle 
  = \frac{\sqrt{\lambda}}{\rho} J_{\nu+\frac{1}{2}(1+\chi)}(\lambda\rho) \ 
  \delta_{\nu\nu'} \delta_{\sigma \sigma'}  \delta_{mm'} \delta_{\chi \chi'} \, ,
\end{equation}
and the completeness relations
\begin{equation}\label{eq:complete}
 {\bf 1} = \sum_{\nu, \sigma, m, \chi} \int_\lambda \left| \lambda; \nu \sigma m \chi \right \rangle 
 \left \langle \lambda; \nu \sigma m \chi \right| 
 = \sum_{\nu, \sigma, m, \chi} \int_\rho \rho^3 
 \left| \rho; \nu \sigma m \chi \right \rangle
 \left \langle \rho; \nu \sigma m \chi \right| \, ,
\end{equation}
where $m$ ranges from 1 to $d^\psi_\nu=\nu(\nu{+}1)$, while $\sigma, \chi =\pm 1$. 
This is all we need to work out the subtraction procedure for the determinant
ratio of Eq.~\eqref{fermdet}.

{\bf Subtraction procedure.}
We proceed like in section~\ref{sec:scalar} and work out
\begin{equation}
\begin{split}
  2 \ln\frac{\det S''}{\det \hat S''} &= 
  \ln \frac{\det \left(-\partial^2 + m_\psi^2 - \cancel{\partial}m_\psi\right)}{
  \det \left(-\partial^2+ \hat{m}_\psi^2\right)} 
  = \tr \ln \left( 1+ \frac{\delta m^2_\psi -\cancel\partial m_\psi}{
  -\partial^2 + \hat m^2_\psi} \right)  
  \\
  &= \tr \frac{1}{-\partial^2}(\delta m^2_\psi - \cancel\partial m_\psi) - 
  \tr \frac{1}{-\partial^2}\hat m_\psi^2 \frac{1}{-\partial^2}
  \left( \delta m^2_\psi - \cancel\partial m_\psi \right)  
  \\ 
  &- \frac{1}{2}\tr \frac{1}{-\partial^2}\left(\delta m^2_\psi - \cancel\partial m_\psi \right) 
  \frac{1}{-\partial^2}\left( \delta m^2_\psi - \cancel\partial m_\psi \right) + \ldots 
  \\
  &= T_{1,0} - T_{1,1} - \frac{1}{2}T_{2,0} + \ldots \, .
\end{split}
\end{equation}
Here $T_{i,j}$ are traces defined like in \eqref{Tracexpand}, where the index $i$ counts
insertions of $\delta m_\psi^2 - \cancel{\partial}m_\psi$, while $j$ counts insertions of 
$\hat m_\psi^2$. 
We have thereby isolated all the three UV divergent traces in $D=4$.
The ellipses stand for terms of higher order in either 
$\delta m_\psi^2 - \cancel{\partial}m_\psi$ or $\hat m_\psi^2$, which are convergent and 
do not need to be regularised.
The rest is the same as in the scalar case; we first subtract the traces evaluated 
in angular momentum space and then add back their dimensionally regularised counterparts
\begin{equation} \label{eq:DetFermi}
  \left.  \ln \frac{\det S^{\prime \prime} }{\det {\hat S}^{\prime \prime} }\right\vert_\psi = 
  \sum_{\nu = 1}^\infty d_\nu^\psi \ln R_\nu - \frac{1}{2}
  \left( T_{1,0} - T_{1,1} - \frac{1}{2} T_{2,0} \right) + \frac{1}{2}
  \left( T^{\rm DR}_{1,0} - T^{\rm DR}_{1,1} - \frac{1}{2} T^{\rm DR}_{2,0} \right) \, .
\end{equation}

Let us start with $T_{1,0}$ and use the same formal manipulations as for 
scalars to get
\begin{align} \nonumber
  T_{1,0} &= \sum_{\nu \sigma m \chi\chi'} \int_\rho \rho^3 
  \int_\lambda  \langle \rho; \nu \sigma m \chi| \frac{1}{-\partial^2}|
  \lambda; \nu \sigma m \chi'\rangle \langle \lambda; \nu \sigma m \chi'|  
  \delta m_\psi^2-\cancel \partial m_\psi | \rho; \nu \sigma m \chi \rangle
  \\ \nonumber
  &= \sum_{\nu \sigma m \chi} \int_\rho \rho^3 \int_\lambda 
  \frac{1}{\lambda^2} \frac{\sqrt{\lambda}}{\rho} J_{\nu+\frac{1}{2}(1+\chi)}(\lambda \rho) 
  \langle \lambda; \nu \sigma m\chi|\delta m_\psi^2|\rho; \nu \sigma m \chi\rangle
  \\ \label{eq:T10_ferm}
  &= \sum_{\nu \sigma m \chi} \int_\rho \rho^3 \delta m_\psi^2 \int_\lambda 
  \frac{1}{\lambda^2} \left(\frac{\sqrt{\lambda}}{\rho} J_{\nu+\frac{1}{2}(1+\chi)}
  (\lambda \rho)\right)^2  
  \\ \nonumber
  &= \sum_{\nu \chi} 2\, d_\nu^\psi \, \frac{1}{2\nu+1+\chi} \int_\rho \rho \ 
  \delta m_\psi^2
  =\sum_\nu d^\psi_\nu \left(\frac{1}{\nu}+\frac{1}{\nu+1}\right) \int_\rho \rho \ 
  \delta m_\psi^2 \, .
\end{align}
In the first line we used the completeness relations to write down the trace in angular
momentum space. 
To go to the second one, we first let $\partial^{-2}$ act on the right to get a bracket 
proportional to $\delta_{\chi\chi'}$.
This implies that the second bracket has to be evaluated only on terms that are diagonal 
in $\chi$ space, such that contributions coming from $\cancel \partial m_\psi$ are 
projected out. 

Next we consider $T_{1,1}$, which is IR-divergent (as for scalars), so we use 
an IR-cutoff regulator $\delir$ to regularise it; it will cancel out later. 
We find
\begin{equation} \label{T11ferm}
\begin{split}
  T_{1,1} &=\hat{m}_\psi^2\sum_{\nu \sigma m \chi} \int_\rho \rho \int_\lambda
  \frac{\lambda}{\left(\lambda^2+\delir^2\right)^2} J^2_{\nu+\frac{1}{2}(1+\chi)}
  (\lambda \rho) \, \delta m_\psi^2
  \\
  &=\hat{m}_\psi^2\int_\rho \rho^3 \delta m_\psi^2 \left(-\frac{1}{12} -\gamma - 
  \ln \frac{\rho \delir}{2} + \sum_{\nu > 1} \frac{d^\psi_\nu}{2 \nu(\nu+1)}
  \left(\frac{1}{\nu-1}+\frac{1}{\nu+2}\right) \right) 
  \\
  &= \hat{m}_\psi^2\int_\rho \rho^3 \delta m_\psi^2 \left(-\frac{1}{12} -\gamma - 
  \ln \frac{\rho \delir}{2} + \frac{1}{2}\sum_{\nu =2}^\infty 
  \left(\frac{1}{\nu-1}+\frac{1}{\nu+2} -\frac{2}{\nu}\right) +\sum_{\nu=2}^\infty \frac{1}{\nu} \right) 
  \\
  &= \hat{m}_\psi^2\int_\rho \rho^3 \delta m_\psi^2 \left( -\gamma - 
  \ln \frac{\rho \delir}{2} -1 +\sum_{\nu=1}^\infty \frac{1}{\nu} \right) \, ,
\end{split}
\end{equation}
where the IR-divergent term comes from $\nu = 1$ and $\chi = -1$.
We used~\eqref{eq:T11_IR} and
$\int_\lambda 1/\lambda^3 J_\nu^2(\lambda\rho) = 
\rho^2/(4\nu \left( \nu^2 - 1 \right))$ for $\nu > 1$. 
The first sum in the third line is equal to $1/12$, and in the last line the sum 
starts at $\nu =1$, compensated by the $-1$ term.
The very last term in \eqref{T11ferm} with the divergent sum encodes the expected
logarithmic UV divergence.

Finally, we calculate $T_{2,0}$. 
After repeated use of the completeness relations~\eqref{eq:complete} and of the scalar
product among $|\rho\rangle$ and $|\lambda\rangle$ states, Eq.~(\ref{eq:rholambda}), 
we get
\begin{equation} \label{eq:T20fer} 
\begin{split} 
  T_{2,0}= &\sum_{\nu \sigma m \chi\chi'} \int_\rho \rho^2 \int_\lambda  
  \lambda^{-\frac{3}{2}}  J_{\nu+\frac{1}{2}(1+\chi)}(\lambda \rho) \int_{\rho'} \rho'^2
  \int_{\lambda'} \lambda'^{-\frac{3}{2}} 
  J_{\nu+\frac{1}{2}(1+ \chi')}(\lambda' \rho') 
  \\
  &\times \langle \lambda; \nu \sigma m \chi| \delta m_\psi^2 - 
  \cancel \partial m_\psi |\rho'; \nu \sigma m  \chi'\rangle \langle \lambda'; 
  \nu \sigma m \chi'| \delta m_\psi^2-\cancel \partial m_\psi |\rho; 
  \nu \sigma m \chi\rangle \, . 
\end{split}
\end{equation}
Here, we let the $\partial^{-2}$ act, while we kept the numerators explicit.
It is the first time that we encounter an expression where the ($\chi$-space) 
off-diagonal term $\cancel \partial m_\psi$ appears twice, so it is not traced out. 
To work out the expression  
we recall that, schematically, 
$\delta m_\psi^2|\rho;\chi\rangle = |\rho;\chi\rangle \,\delta m_\psi^2(\rho)$, while 
$\cancel \partial m_\psi|\rho;\chi\rangle = |\rho;-\chi\rangle \,\dot m_\psi(\rho)$ 
induces a flip of chirality (cf. Eq.~\eqref{blockDdagD}) and $T_{2,0}$ finally reduces to
\begin{equation}\label{eq:T20_pre}
  T_{2,0} = \sum_{\nu = 1}^{\infty} d_\nu^\psi \int_\rho \rho^3
  \int_0^1 {\rm d} x \left[ \bigg(\frac{x^{2\nu+1}}{\nu^2} + \frac{x^{2\nu+3}}
  {(\nu + 1)^2}\bigg) \delta m_\psi^2 \delta m_\psi^2(x\rho) +
  \frac{2 x^{2\nu+2}}{\nu(\nu + 1)}  \dot{m}_\psi \dot{m}_\psi(x\rho ) \right] \, .
\end{equation}
This expression is analogous to~\eqref{T20ang} in the scalar sector, although 
more complicated, and the obvious next step is to replicate the main lesson 
we have learned there.
We split $\delta m_\psi^2(x\rho) = \delta m_\psi^2 + (\delta m_\psi^2(x\rho) - \delta m_\psi^2)$
and $\dot m_\psi(x\rho) = \dot m_\psi +(\dot m_\psi(x\rho) - \dot m_\psi)$. 
Doing so we get terms with $(\delta m_\psi^2)^2$ and $\dot m^2_\psi$ for which the integral
in $x$ can be trivially performed, plus terms with $(\delta m_\psi^2(x\rho) - \delta m_\psi^2)$ 
and $(\dot m_\psi(x\rho) - \dot m_\psi)$ for which the $\sum_\nu$ is convergent and can be performed. 
This gives
\begin{align} 
\begin{split}
  T_{2,0} &= \frac{1}{2} \sum_{\nu=1}^\infty \left(\frac{1}{\nu} + 
  \frac{\nu}{(\nu+1)(\nu+2)} \right) \int_\rho \rho^3 (\delta m_\psi^2)^2
  + 2 \sum_{\nu=1}^\infty \frac{1}{2\nu +3} \int_\rho \rho^3 \dot m^2_\psi
  \\
  & + \int_\rho \rho^3 \delta m_\psi^2 \int_0^1 {\rm d}x \frac{2x^3}{1-x^2}
  \left( \delta m_\psi^2(x\rho) - \delta m_\psi^2 \right)  
  \\ 
  & + \int_\rho \rho^3 \dot m_\psi \int_0^1 {\rm d}x \frac{2x^4}{1-x^2} 
  \left( \dot m_\psi(x\rho) - \dot m_\psi \right) 
\end{split}
\\ \label{T20ferm}
\begin{split}
  &= \sum_{\nu=1}^\infty \frac{1}{\nu}\int_\rho \rho^3 \left((\delta m_\psi^2)^2  
  + \dot m_\psi^2 \right) - \int_\rho \rho^3 (\delta m_\psi^2)^2
  + \int_\rho \rho^3 \dot m^2_\psi \left(-\frac{8}{3} + \ln 4 \right)
  \\
  & + 2\int_\rho \rho^3 \int_0^1  \frac{{\rm d}x\,x^3}{1-x^2} \bigg[\delta m_\psi^2 
  \left(\delta m_\psi^2(x\rho) - \delta m_\psi^2 \right) +
  x \,\dot m_\psi \left( \dot m_\psi(x \rho) - \dot m_\psi \right)\bigg] \, .
\end{split}
\end{align}
\sloppy To get to the second equality we added and subtracted $1/\nu$ in the first sum, then used
$\sum_{\nu=1}^\infty \left( \nu/(\nu+1)/(\nu+2) - 1/\nu \right) = -2$. 
In the second sum we added and subtracted $1/(2\nu)$ and used
$\sum_{\nu = 1}^\infty \left( 1/(2 \nu + 3) - 1/(2\nu) \right) = 
-4/3 + \ln 2$.
We have again isolated the logarithmic divergence in the first term 
of~\eqref{T20ferm}, and found several other finite pieces. 
These will combine with those coming from the dimensionally regularised traces, as
we shall see.

Our next task is to evaluate the traces $T_{i,j}^{\rm DR}$. 
This can be done, {\it mutatis mutandis}, with the same method we used in the scalar sector, 
together with some standard Dirac-matrix algebra. 
Here we quote directly the results for the fermionic case:
\begin{align} 
  T^{\rm DR}_{1,0} &= 0 \, ,
  \\
  T^{\rm DR}_{1,1} &= \hat{m}^2_\psi \left(\frac{1}{\epsilon}+
  \ln \frac{\tilde \mu}{\delir}\right) \int_\rho \, \rho^3\,\delta m^2_\psi \, ,
  \\
  T^{\rm DR}_{2,0}&= \frac{1}{4\pi^2} \int_k \left(\big|\widetilde{\delta m^2_\psi}(k) 
  \big|^2 + k^2 \big|\widetilde{\, m_\psi}(k)\big|^2\right) 
  \left( \frac{2}{\epsilon} + 
  2 + \ln \frac{\tilde\mu^2}{k^2} \right) \, . \label{T20DRferm}
\end{align}
We can put $T_{2,0}^{\rm DR}$ in a more useful form by massaging 
the $k$ integrals with $\ln k$. 
We do this by using~\eqref{eq:Fourier} for the first term proportional to $\widetilde{\delta m^2_\psi}(k)$. 
For the second term with an additional factor of $k^2$ coming from the Fourier
transform of $\cancel \partial\cancel \partial$, we employ the following identity,
derived in Appendix~\ref{app:FTProofs},
\begin{equation}\label{eq:Fourier2}
  \kern-1.5em \int_k \left|\tilde{f}(k) \right|^2 k^2 \ln k^2 = 
  -4 \pi^2 \int_\rho \rho^3{\dot f}(\rho)\left[ {\dot f}(\rho) 
  \left( \gamma_E{-}\frac{8}{3}{+}\ln ({2\rho}) \right) + 2 \int_0^1 {\rm d} x\, 
  x^4\, \frac{{\dot f}(x \rho){-}{\dot f}(\rho)}{1-x^2} \,\right] ,
\end{equation}
where $\tilde f(k)$ is the Fourier transform of the radial function $f(\rho)$.
We obtain
\begin{equation} \label{T20DRfermfin}
\begin{split}
  T_{2,0}^{\rm DR} &= \int_\rho \rho^3 (\delta m_\psi^2)^2 \left( \frac{1}{\epsilon} + 
  \gamma_E + \ln \frac{\tilde \mu \rho}{2} \right) + 
  \int_\rho \rho^3 \dot m_\psi^2 \left( \frac{1}{\epsilon} + \gamma_E - \frac{5}{3} + 
  \ln(2\tilde \mu \rho)  \right) 
  \\
  & + \int_\rho \rho^3 \delta m_\psi^2 \int_0^1 {\rm d}x \frac{2x^3}{1-x^2}
  \left( \delta m_\psi^2(x\rho) - \delta m_\psi^2 \right)  
  \\ 
  & + \int_\rho \rho^3 \dot m_\psi \int_0^1 {\rm d} x \frac{2x^4}{1-x^2} 
  \left(\dot m_\psi(x\rho) - \dot m_\psi \right) \, .
\end{split}
\end{equation}

Combining all the ingredients together in~\eqref{eq:DetFermi}, the last two lines 
in~\eqref{T20ferm} cancel against the ones in~\eqref{T20DRfermfin} and the
double integrals disappear.
After a little algebra to combine the remaining parts we find
\begin{equation} \label{eq:ferm_final}
\begin{split}
 ~~\ln \frac{{\det} S''}{{\det} {\hat S''} } 
  &= \sum_{\nu = 1}^\infty \left( d_\nu^\psi \, \ln R_\nu^\psi - \left(\nu+\tfrac12\right) 
  \int_\rho \, \rho \, \delta m_\psi^2 
 + \frac{1}{4 \nu} \int_\rho \, \rho^3 \left( \delta m_\psi^4+{\dot{m}_\psi}^2 \right) \right)
 \\
 &- \frac{1}{4}  \int_\rho \, \rho^3 \left( \delta m_\psi^4+{\dot{m}_\psi}^2 \right) 
  \left( \frac{1}{\epsilon} + 1 + \gamma_E + \ln \frac{\tilde \mu \rho}{2} \right) \, ,
\end{split}
\end{equation}
which matches the form of \eqref{eq:DetS_ferm}.

{\bf Generalisation to many Majorana particles.} 
We can generalise the result in many ways. 
We proceed in three steps of increasing difficulty. 
We could first consider the theory of a single Dirac spinor and a scalar $\phi$ with 
a parity violating Lagrangian term
\begin{equation}\label{eq:gamma5}
  {\rm Re}(m+y\phi) \ \bar{\psi}\psi - i\,{\rm Im}(m+y\phi) \ \bar{\psi} \gamma_5 \psi \, ,
\end{equation}
with $m$ and $y$ complex. 
Calling $m_\psi(\phi) = m + y \phi$, we see that the (squared) differential 
operator $S''^\dagger \!S''$ on a $\phi$ background becomes
\begin{equation}
  S''^\dagger \!S'' = -\partial^2 + \left| m_\psi(\phi) \right|^2 -
  \cancel \partial \left(\frac{1-\gamma_5}{2} m_\psi(\phi) + 
  \frac{1 + \gamma_5}{2} m_\psi(\phi)^* \right) \, ,
\end{equation}
and the partial $\nu$-operator \eqref{blockDdagD} gets promoted to
\begin{equation} 
  \left(S''^\dag \!S''\right)^{(\nu)}=
  \begin{pmatrix} 
  - \Delta_{\nu-1}+|m_\psi|^2    &  -\dot m_\psi 
  \\ 
  - (\dot m_\psi)^*         &  -\Delta_{\nu}+|m_\psi|^2
  \end{pmatrix} \, .
\end{equation}
To go to the previous case, one just has to take $m,y$ real. 
On the contrary, to promote~\eqref{eq:ferm_final} it is enough to put an absolute
value $\| \cdot \|$ sign whenever necessary.

The second obvious generalisation is to many Dirac particles and a parity violating 
interaction like in \eqref{eq:gamma5}. 
Here the same comments we made at the end of Sec.~\ref{sec:scalar} apply, and we 
essentially just need to add indices and put `tr' in the final 
expression~\eqref{eq:ferm_final}, because now $\delta m_\psi^2$ and friends are matrices 
in the flavour indices. 
We only quote this time the expression for the partial $\nu$-operator, which has now 
also flavour indices
\begin{equation} 
 \left(S''^\dag \!S''\right)_{ij}^{(\nu)} = \begin{pmatrix} 
 -\delta_{ij}\Delta_{\nu-1}+\big(m_\psi^\dagger m_\psi\big)_{ij} &  
 -\big(\dot m_\psi\big)_{ij} 
 \\ 
 - \big(\dot m_\psi^\dagger\big)_{ij} &   
 -\delta_{ij}\Delta_\nu+\big(m_\psi^\dagger m_\psi\big)_{ij} 
 \end{pmatrix}\,.
\end{equation}
It can be pictured as a square matrix of dimension $2N_f$, with $N_f$ the number of flavours.

\vspace{.3cm}
 
There is a third possible generalisation. 
Like a theory for $N_s$ complex scalars --- invariant under rephasing of each scalar --- 
is not the same as a theory of $2N_s$ real scalars, but is more restrictive since the latter 
allows $U(1)^{N_s}$ violating terms, a theory of $N_f$ Dirac fermions is more restrictive 
than a theory of $2 N_f$ Majorana fermions.

Before exploring the generalisation to a theory with Majorana-type particles, let us 
recall what happens in the scalar analogue, keeping in mind that our goal is to compute 
functional determinants of fluctuation operators. 
Schematically, in the complex case the determinant comes from a path integral like
$\int {\cal D}\phi\,{\cal D}\bar\phi \exp \left[
{\bar\phi(-\partial^2+m^2)\phi}\right]\sim 
\det(-\partial^2+m^2)^{-1}$,
where $m^2$ is a hermitian matrix (we suppress flavour indices), while in the real case one 
has a path integral like
$\int {\cal D}\phi \exp\left[{\frac{1}{2}\phi(-\partial^2+m^2)\phi}
\right]\sim \det(-\partial^2+m^2)^{-\frac{1}{2}}$,
with $m^2$ being real and symmetric. 
The difference among the two instances is that there is an extra 
factor of $\frac{1}{2}$ in the action of the real scalar, due to canonical
normalisation of the kinetic and mass terms, in addition to an overall 
square root in the RHS of the expression. 
Notice that, keeping the number of degrees of freedom fixed, the second case 
is more general, since a hermitian $N_s\times N_s$ matrix has less parameters 
than a real and symmetric $(2N_s)\times(2 N_s)$ matrix 
($N_s^2$ versus $2N_s^2+N_s$). 
Despite these differences, the most important lesson is that we end up
with a formally equal determinant.

Coming back to fermions, in the Majorana basis (of gamma matrices) a Majorana
spinor is just a Dirac spinor $\xi_\alpha(x)$ with real entries.
Due to canonical normalisation, its Lagrangian will be formally the same as 
that of a Dirac particle, apart from factors of $\frac{1}{2}$ in the 
kinetic and mass terms~\cite{Pal:2010ih}. 
The path integral then reads 
$\int {\cal D}\xi \exp\left[\frac{1}{2}\bar\xi(i\cancel\partial + m)
\xi \right]\sim {\rm Pf}(i\cancel\partial +m)$, 
where `Pf' denotes the Pfaffian of the operator, which is equal to the square
root of its determinant up to a sign~\cite{Zinn-Justin:572813}. 
Since we are interested in determinant~\emph{ratios} among an operator and 
its free theory limit, signs cancel in the ratio due to continuity in, say, 
the $y\to 0$ limit.
In conclusion, when dealing with Majorana species, 
$\det(\cdot) \to \sqrt{\det(\cdot)}$. 
In practice,~\eqref{eq:DetS_ferm} is unchanged, while~\eqref{eq:master} has 
to include a square root also for the fermionic determinant ratio. 
Reality conditions on the mass matrix are also to be taken into account, 
like in the scalar case.

Let us give an example of the use of~\eqref{eq:DetS_ferm} in presence of 
two Majorana particles, with fields $\eta$ and $\xi$. 
Let the mass term be
\begin{equation}
  L_m = \frac12 \left( m_\eta \bar \eta \eta + m_\xi \bar\xi \xi + 
  m_D \left( \bar\eta \xi + \bar\xi \eta\right) + i m_\eta^5 
  \bar\eta \gamma_5\eta + i m_\xi^5 \bar\xi \gamma_5\xi + i m_D^5\left(\bar \eta 
  \gamma_5\xi+\bar\xi \gamma_5\eta\right)\right),
\end{equation}
where, in the presence of a bounce or a field VEV, the masses include also the 
background contribution.
The $M_{ab}$ that enters into~\eqref{eq:mpsidef} is then given by
\begin{equation} \label{eq:M22}
  M = \begin{pmatrix}
    m_\eta + i m_\eta^5 & m_D + i m_D^5 
    \\
    m_D + i m_D^5       & m_\xi+i m_\xi^5
    \end{pmatrix} \, ,
\end{equation}
which is a $2\times 2$ complex symmetric matrix. 
If $m_{\eta,\xi}=m_{\eta,\xi}^5=0$ the two Majorana degrees of freedom allow in 
principle a Dirac structure. 
In this limit one has two possibilities. 
The first is to keep a $2 \times 2$ matrix like in~\eqref{eq:M22} and at the same 
time add a square root to the fermion determinant in~\eqref{eq:master}. 
The second is to keep~\eqref{eq:master} as it is and consider a $1{\times}1$ mass 
matrix $M=m_D+im_D^5$. 
The two approaches give the same result.

%
%
\section{Vectors} \label{sec:vector}
In this section we build the regularised functional determinant for gauge boson fluctuations. 
We consider only the Abelian case. As explained shortly after~\eqref{eq:gauge_inv}, 
the generalisation to the non-Abelian case is simple. 

We start with the Euclidean action for a $U(1)$ gauge field $A_\mu$, coupled to a charged 
scalar $\Phi=(h + i a)/\sqrt{2}$, as given in~\eqref{eq:Lgauge}.
Recall that we work in the background $R_\xi$ gauge and fix $\xi=1$.
Expanding the action to second order in the fields around the bounce, we get $\int_xL^{(2)}$, 
with
\begin{equation}
  L^{(2)} = \frac12 h \left( -\partial^2 + m_h^2 \right)h +
  \frac12 {\cal A}_I \, S''_{IJ} {\cal A}_J + 
  \bar c \left(-\partial^2 + m_A^2 \right) c \, ,
\end{equation}
where ${\cal A}_I = (A_\mu, a)$, and the fluctuation operator $S''$ is now a $5 \times 5$ 
matrix
\begin{equation}\label{eq:Aa}
  S''  = \begin{pmatrix}
  (-\partial^2+m_A^2)\,\delta_{\mu\nu}&2\,\dot m_A\, \hat x_\mu
  \\
  2\,\dot m_A\, \hat x_\nu &-\partial^2 + m_a^2
  \end{pmatrix} \, .
\end{equation}
The $\rho$-dependent masses $m_A^2$, $\dot m_A$ and $m_a^2$ are defined in~\eqref{eq:maA}, 
and $\hat x_\mu$ comes from $\partial_\mu \bar\phi = \hat x_\mu\partial_\rho\bar\phi$. 
The ghosts $c,\bar c$ and the physical mode $h$ are decoupled from the other degrees 
of freedom. 
Given that they are scalars, we refer to Section~\ref{sec:scalar} for their subtraction 
prescription, and focus on $A_\mu$ and $a$ here.

{\bf Multipole decomposition.} We want to compute the determinant of $S''$. 
Like in the previous sections, we are going to exploit the $O(4)$ symmetry of the 
action \textit{and} the bounce solution, which guarantees that $[J_{\mu\nu}, S'']=0$, 
and look for eigenstates of the fluctuation matrix with definite total angular momentum 
quantum numbers. 

\begin{figure}[t]
  \centering
  \begin{tikzpicture}[line width=1.1 pt, scale=.7, baseline=(current bounding box.center)]
		\draw[-stealth] (0,0) to  (0,5.5) ;
		\draw[-stealth] (0,0) to  (5.5,0) ;
		
		\node at (0,6) {$j_B$};
		\node at (6,0) {$j_A$};
				
		\foreach \x in {0,1,2}
		{\node at (2*\x,-.5) {\footnotesize $\x$} ;
		\node at (-.7,2*\x) {\footnotesize $\x$} ;}
		\foreach \x in {1,3,5}
		{\node at (\x,-.5) {\footnotesize $\frac{\x}{2}$} ;
		\node at (-.7,\x) {\footnotesize $\frac{\x}{2}$} ;}
		
		\filldraw[black] (0,0) circle (.5pt) ;
		\filldraw[black] (0,2) circle (.5pt) ;
		\filldraw[black] (2,0) circle (.5pt) ;
		\filldraw[black] (4,6) circle (.5pt) ;
		\filldraw[black] (6,4) circle (.5pt) ;

  		\foreach \x in {1,2,3,4,5} \foreach \y in {1,2,3,4,5}
  		\filldraw[black] (\x,\y) circle (.5pt) ;
		
		\foreach \x in {0,1,2,3,4,5}
    		\draw[black!10!cyan,line width =.8] (\x,\x) circle (5pt) ;
		\foreach \x in {1,2,3,4,5}
		\draw[black!50!cyan,line width =.8] (\x,\x) circle (9pt) ;

		\foreach \x in {0,1,2,3,4}
    		{\draw[olive,line width =.8] (\x,\x+2) circle (5pt) ;
		\draw[olive,line width =.8] (\x+2,\x) circle (5pt) ;
		}
		

		
	 \end{tikzpicture}
  \vspace{.2cm}
  \caption{$SO(4)$ multiplet decomposition of a vector field $A_\mu(\hat x)$.
  The modes on the diagonal are called $D^\pm$ and come in pairs with orbital
  angular momentum $l=j_A+j_B\pm 1$. The modes off the diagonal are called
  $T^\pm$ and have orbital angular momentum $l=j_A+j_B$.}
  \label{fig:vectors}
\end{figure}
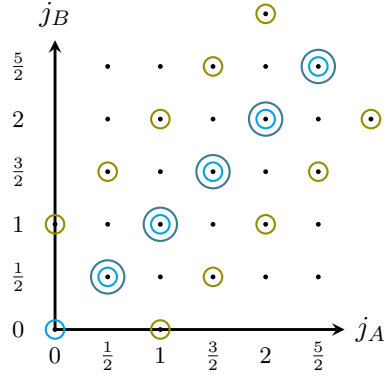

The would be NGB $a(x)$ is a scalar, so we expand it on a basis of spherical harmonics 
like in Sec. \ref{sec:scalar}. 
On the other hand, from the point of view of rotations a gauge field $A_\mu(x)$ is a 
tensor product among a scalar in the $x$ variable and the $(1/2, 1/2)$ 
representation of $SO(4)$ in the $\mu$ index. 
Carrying out the tensor product among the two, similarly to what we did for fermions 
in Sec.~\ref{sec:fermion}, we get
\begin{equation}\label{eq:group_th_vector}
  \bigoplus_{l=0}^\infty \,(\tfrac{1}{2},\tfrac12 )\otimes (\tfrac{l}{2},\tfrac{l}{2}) 
  \simeq \underbrace{(0,0)_1}_{D_0^+}\oplus \bigoplus_{j=1}^\infty \left(\underbrace{(\tfrac{j}{2},\tfrac{j}{2})_{j-1}}_{D^-_j}
  \oplus \underbrace{(\tfrac{j}{2},\tfrac{j}{2})_{j+1}}_{D^+_j} 
  \oplus \underbrace{(\tfrac{j-1}{2},\tfrac{j+1}{2})_j}_{T^-_j}
  \oplus \underbrace{(\tfrac{j+1}{2},\tfrac{j-1}{2})_j}_{T^+_j} \right) \, ,
\end{equation}
where the suffix $l$ in $(j_A,j_B)_l$ labels the eigenvalue of the $SO(4)$ multiplet 
under $L^2$, equal to $l(l+2)$.
Following \cite{Baratella:2024hju}, we call $D_j^\pm$ the diagonal multiplets 
$(j/2, j/2)_{j \pm 1}$, and $T_j^\pm$ the transverse multiplets 
$((j \pm 1)/2,(j \mp 1)/2))_j$. 
This is depicted in Fig.~\ref{fig:vectors}.
We can then expand the generic gauge field into these modes as
\begin{equation}\label{eq:ang_Amu}
  A_\mu(\hat x,\rho) = \sum_{{\cal I},\pm} \left(
  \alpha^\pm_{\cal I}(\rho) \,  D^\pm_{\mu,\cal I}(\hat x) +
  \beta^\pm_{\cal I}(\rho)  \,  T^\pm_{\mu,\cal I}(\hat x) \right) \, ,
\end{equation}
where the multi-index ${\cal I}=\{j,m_A,m_B\}$ runs over all the quantum numbers 
that are needed to specify the angular modes.
For example, for $D^+$ we have $j= 0, 1, 2, \ldots$ and as usual 
$|m_{A,B}| \leq j/2$ run in integer steps.
The $\alpha_{\cal I}^\pm$ and $\beta_{\cal I}^\pm$ are radial functions that fully 
characterise the given configuration $A_\mu(x)$. 
For the explicit form of the angular modes $D^\pm_{\mu,\cal I}(\hat x)$ 
and $T^\pm_{\mu,\cal I}(\hat x)$ we refer to~\cite{Baratella:2024hju}.
Here we only need to know their group theoretical properties.

With the field decomposition defined in~\eqref{eq:ang_Amu}, we let $S''$ act on the 
$(A_{\mu},a)$ vector.
Thanks to rotational invariance, we can anticipate the form that the partial $j$-operators
will take.
The $D^\pm$ modes have the same transformation properties under rotations as scalars,
so under the action of $S''$ they can mix with one another and with the $a$ modes. 
Conversely, there is no other mixing allowed, so in particular the $T^\pm$ modes are 
decoupled from each other and from $\{ D^\pm, a \}$. 
Therefore, the determinant of the fluctuation matrix will be of the form
\begin{align} \label{eq:gauge_decomp}
  \det S''&= \left[\ 
  \prod_{j=1}^\infty\left(\det{ S''_j}^{(T)}\right)^{j(j+2)}\right]^{2} 
  \ \  
  \prod_{j=0}^\infty\left(\det{S''_j}^{(Da)}\right)^{(j+1)^2} \, ,
\end{align}
where the power of 2 outside of the square brackets accounts for the two transverse 
modes $T^+$ and $T^-$, whose fluctuation operator is identical due to parity, while the 
power $j(j+2)$ accounts for their degeneracy. 
Similarly, the power of $(j+1)^2$ accounts for the degeneracy of diagonal modes and 
scalars.\footnote{The degeneracy factor equals the dimension of the representation, 
which is $(2j_A+1)(2j_B+1)$ for a $(j_A,j_B)$ representation.}

To calculate the explicit form of partial operators, we are going to exploit the 
rotational properties of our angular basis. 
In particular, we know on general grounds that
$\partial^2\big[c(\rho) {\cal B}_\ell(\hat x)\big] = 
{\cal B}_\ell(\hat x)\Delta_\ell \big[c(\rho)\big]$, 
where $\Delta_\ell$ is given in~\eqref{eq:def_Dj} and ${\cal B}_\ell$ is any angular 
function with definite orbital angular momentum, such that 
$L^2 {\cal B}_\ell = \ell(\ell+2){\cal B}_\ell$. 
The basis in~\eqref{eq:ang_Amu} has definite orbital angular momentum by construction, 
so the action of $S''$ in~\eqref{eq:Aa} is mostly diagonal on it, except for
\begin{equation}
  S''_{\rm o.d.} \equiv 2\, \dot m_A \,
  \begin{pmatrix}
    0           & \hat x_\mu \\
    \hat x_\nu  & 0
  \end{pmatrix} \, .
\end{equation}
To understand the action of $S''_{\rm o.d.}$ on the modes, it is useful to first decompose
$A_\mu(x) = A_\mu^\perp(x) + A^r(x)\hat x_\mu$, where $A_\mu^\perp \hat x_\mu = 0$.
Then we find
\begin{align}
  S''_{\rm o.d.}\begin{pmatrix}
    A \\ a
  \end{pmatrix} &= 2 \, \dot m_A \begin{pmatrix}
    a\, \hat x \\ A^r
  \end{pmatrix}, 
  & \left( S''_{\rm o.d.} \right)^2 \begin{pmatrix}
    A \\ a
  \end{pmatrix} &= 4 \, \dot m_A^2 \begin{pmatrix}
    A^r\, \hat x \\ a
  \end{pmatrix} \, ,
\end{align}
which shows that, up to the overall factor of $4\,\dot m_A^2$, the $(S''_{\rm o.d.})^2$ 
acts as a projector on the space of outward pointing vector fields, and as the identity 
on the $a$ space. 
The effect of $S''_{\rm o.d.}$ is to annihilate $A^\perp_\mu$ and swap $a$ and $A^r$. 
When we operate with $S''_{\rm o.d.}$ on a basis element for the field $a(x)$, that is 
on spherical harmonics, and given that $[J_{\mu\nu},S''_{\rm o.d.}] = 0$, we have
\begin{equation}\label{eq:Mod}
  \frac1{2\,\dot m_A}\,S''_{\rm o.d.} Y_{jm_Am_B}(\hat x)=\hat x\, Y_{jm_Am_B}(\hat x) =
  c^+ D^+_{\mu,\,jm_Am_B}(\hat x) + c^- D^-_{\mu,\,jm_Am_B}(\hat x) \, .
\end{equation}
The effect of $S''_{\rm o.d.}$ on $Y_{jm_Am_B}$ must be a linear combination of $D^\pm$ 
with the same $j,m_A,m_B$ indices as the spherical harmonic. 
The linear combination that relates the outward-pointing modes and the definite-$L^2$ 
ones is given in~\cite{Isidori:2001bm}, and sets 
\begin{align}
  c^\pm &= \pm \sqrt{\frac{j + 1 \pm 1}{2(j+1)}} \, .
\end{align}
We can now write down the partial $j$-fluctuation matrices
\begin{align} \label{eq:gauge_fluc_Ba}
  {S''}_0^{(Da)} &= \begin{pmatrix}
    -\Delta_1 + m_A^2 & 2 \, \dot m_A
    \\
    2\,\dot m_A&-\Delta_0+m_a^2
  \end{pmatrix}, 
  \\ \label{eq:gauge_fluc_BLa}
  {S''}^{(Da)}_{j>0} &=
  \begin{pmatrix}
    -\Delta_{j-1}+m_A^2     & 0 &  -2\,\dot m_A\sqrt{\frac{j}{2(j+1)}}
    \\
    0 &-\Delta_{j+1}+m_A^2  & 2\,\dot m_A\sqrt{\frac{j+2}{2(j+1)}}   
    \\
      -2 \,\dot m_A\sqrt{\frac{j}{2(j+1)}} & 2\,\dot m_A \sqrt{\frac{j+2}{2(j+1)}}
    &-\Delta_j + m_a^2
  \end{pmatrix} \,,
  \\ \label{eq:gauge_fluc_T}
  {S''}^{(T)}_{j>0} &= -\Delta_j + m_A^2 \, ,
\end{align}
that were first derived in~\cite{Isidori:2001bm} in the $\xi=1$ background gauge 
that we are using here.
Note that~\eqref{eq:gauge_fluc_Ba} has only two entries, because there is no $SO(4)$ 
singlet $D^-$ mode. 
In order to derive the partial operators $\hat {S}''_j$ on the false vacuum background, 
one just needs to substitute $m_i \to \hat m_i$ everywhere and obtain operators that are 
all diagonal in our ``$L^2$-basis''.

{\bf Hilbert space structure.} The structure of the Hilbert space of fluctuations is 
remarkably more complex here than in the previous sections.
Initially, we thought of the Hilbert space where~\eqref{eq:Aa} operates as 
${\cal H}={\cal H}_A\oplus{\cal H}_a$, i.e. as the Cartesian product of $A_\mu(x)$ 
and $a(x)$ fluctuations.
Following group-theoretical arguments, we were able to split 
${\cal H}_A={\cal H}_{D^-}\oplus{\cal H}_{D^+}\oplus{\cal H}_{T^-}\oplus{\cal H}_{T^+}$, 
where in turn each element is further split into subspaces with definite total angular momentum, 
for example ${\cal H}_{D^-}=\bigoplus_{j,m_A,m_B}{\cal H}_{D^-}^{jm_Am_B}$. 
Each such subspace is still infinitely-dimensional: the objects in ${\cal H}_{D^-}^{jm_Am_B}$ are 
functions of the form $c(\rho) D^-_{\mu,jm_Am_B}(\hat x)$, where $c(\rho)$ is arbitrary up to 
regularity conditions. 
We have learned that, modulo the effect of $S''_{\rm o.d.}$, the fluctuation operator 
of~\eqref{eq:Aa} acts diagonally on each of these subspaces. 
The only effect that $S''_{\rm o.d.}$ can have is to jump from ${\cal H}_{a}^{jm_Am_B}$ to 
${\cal H}_{D^-}^{jm_Am_B}\oplus{\cal H}_{D^+}^{jm_Am_B}$ and back.
We are going to make use of this structure to compute traces in what follows.
In particular, first we will work in the {\em smallest possible Hilbert subspace} with definite 
total angular momentum, and then sum up the contributions from all modes $a,D^\pm,T^\pm$ and 
all angular momentum quantum numbers $j, m_A, m_B$. 
Working in the smallest subspaces has the practical advantage of not complicating our formulas 
with notation like $|D^+,jm_Am_B;\rho\rangle$, or `${\rm tr}_{T^-jm_Am_B}$'.
We simply use $\ket \rho$ for the state localised in $\rho$-coordinate and normalised as 
$\langle \rho| \rho'\rangle = \rho^{-3}\delta(\rho-\rho')$, and
$|\lambda\rangle$ for the vector in ${\cal H}_{\alpha}^{jm_Am_B}$ that satisfies 
$\partial^2|\lambda\rangle = \Delta_{\ell_{\alpha , j}}|\lambda\rangle = -\lambda^2|\lambda\rangle$ 
and is normalised as $\langle \lambda|\lambda'\rangle=\delta(\lambda-\lambda')$.
We also have $\langle \rho|\lambda\rangle=\rho^{-1}\lambda^{\frac12}J_{\ell_{\alpha,j}+1}(\lambda\rho)$, 
with $J$ the Bessel function.
We recall that $\Delta_\ell$ is the operator given in~\eqref{eq:def_Dj}. 
With $\ell_{\alpha,j}$ we refer to the orbital angular momentum of a state 
$\alpha \in \{ a, D^\pm, T^\pm \}$, coming from the multipole decomposition~\eqref{eq:group_th_vector}.

{\bf Subtraction procedure.}
We can now work out the regularisation of the functional determinant for gauge bosons. 
Following the same procedure as for scalars and fermions, we isolate the divergent traces $T_{1,0}$,
$T_{1,1}$ and $T_{2,0}$ and compute them in the angular momentum and in the momentum basis using DR. 
Using~\eqref{eq:gauge_decomp} and performing the same formal steps as in~\eqref{lndettrace}, 
we get to the following sum over the various sectors
\begin{align}\label{eq:exp_vec}
  \!\!\! \ln\frac{\det{S''}}{\det\hat{S}''} =\sum_{\sigma,j} d_j^\sigma \,{\rm tr} \,
  \ln \left[ 1 + {\left( {\hat S}^{\prime \prime (\sigma)}_j \right)}^{-1}
  \delta {S''}_j^{(\sigma)} \right] =
  \sum_{\sigma,j} d_j^\sigma \left(
  T_{1,0}^{\sigma,j}-T_{1,1}^{\sigma,j} - \frac12 T_{2,0}^{\sigma,j} \right) + \ldots \, ,
\end{align}
where the ellipses stand for terms that are not UV divergent.
We have defined $\delta {S''} = S'' - \hat S''$ and $\hat S'' = (-\partial^2+\hat m^2)$.
In the second step we have expanded both the logarithm around {\bf 1} and 
$\hat S''$ in powers of $\hat m^2$.
As in previous sections we use the notation $T_{I,J}$, with the index $I$ counting insertions 
of $\delta {S''}$ and $J$ insertions of $\hat m^2$.

Following the discussion above, we compute the traces in subspaces defined by $j$ and sectors
$\sigma = T^+, T^-, Da$. 
As usual $m_A$ and $m_B$ are sterile indices, so we do not carry them around, but take into 
account their multiplicity in the degeneracy factors
\begin{align} \label{eq:djvect}
  d_j^{Da}      &\equiv d_j^\phi = \left( j + 1 \right)^2 \, , &    
  d_j^{T^\pm}   &\equiv d_j^T = j \left( j + 2 \right) \, .
\end{align}  
 
By inspection of~\eqref{eq:exp_vec} we note that ${\hat S}^{\prime \prime (\sigma)}_j$ is diagonal 
in each space ${\cal H}_\alpha^{jm_Am_B}$.
The same holds for $\delta {S''}_j^{(\sigma)}$, except for the off-diagonal effects of 
$S''_{\rm o.d.}$ that we described earlier. 
The $S''_{\rm o.d.}$ only affects the $T_{2,0}$ sector, because a purely off-diagonal matrix $x$ 
has ${\rm tr}[xd]=0$ for any diagonal matrix $d$, and one needs at least a second power of $x$ 
in order not to have a vanishing trace.
It is analogous to what we encountered in the fermion case in Sec.~\ref{sec:fermion}, where 
the off-diagonal part of the matrix entered only in the calculation of $T_{2,0}$.
In light of this, it is natural to split
\begin{align}
  T_{1,J}^{Da,j} &= T_{1,J}^{a,j} + T_{1,J}^{D^+,j} + T_{1,J}^{D^-,j}\,, 
  & 
  T_{2,0}^{Da,j} &= T_{2,0}^{a,j} + T_{2,0}^{D^+,j} + T_{2,0}^{D^-,j}+T^j_{\rm o.d.} \,,
\end{align}
where $T^j_{\rm o.d.}$ is the only trace where the off-diagonal element enters.

Let us start with $T_{1,0}$. We have
\begin{align}
\begin{split}
    T_{1,0}^{\alpha,j} &= {\rm tr}\left(\frac{1}{-\partial^2}\delta m_\alpha^2\right)
    = -\int_\rho \rho^3\int_\lambda\langle\rho|\Delta_{\ell_{\alpha,j}}^{-1} |
    \lambda\rangle\langle\lambda|\delta m_\alpha^2 |\rho\rangle \nonumber 
    \\
    &= \int_\rho \int_\lambda \frac{\rho}{\lambda}\delta m_\alpha^2 (\rho)
    J_{\ell_{\alpha,j} + 1}(\lambda\rho)^2
    = \frac1{2(\ell_{\alpha,j} + 1)} \int_\rho \rho\,\delta m_\alpha^2 \, .
\end{split}    
\end{align}
The index $\alpha$ runs over all species $a,D^\pm,T^\pm$, while $\ell_{\alpha,j}$ label here is related to orbital angular momentum and can be read off 
from~\eqref{eq:group_th_vector}
\begin{align} \label{eq:orbalpha}
  \ell_{a,j} &= \ell_{T^\pm,j} = j \, , & \ell_{D^\pm,j} &= j \pm 1 \, .
\end{align}
With similar manipulations, for $T_{1,1}$ we get
\begin{equation}
\begin{split}
  T_{1,1}^{\alpha,j} &= {\rm tr}\left( \frac{1}{-\partial^2}\hat m_\alpha^2 
  \frac{1}{-\partial^2}\delta m_\alpha^2 \right)
  \\
  &= \hat m_\alpha^2\int_\rho\int_\lambda \frac{\rho}{\lambda^3}
  \delta m_\alpha^2(\rho)J_{\ell_{\alpha,j}+1}(\lambda\rho)^2
  = \frac{1}{4\,\ell_{\alpha,j}(\ell_{\alpha,j}+1)(\ell_{\alpha,j}+2)} 
  \int_\rho \rho^3 \hat m_\alpha^2 \delta m_\alpha^2 \, .
\end{split}
\end{equation}
Due to the presence of the $\ell_{\alpha,j}^{-1}$ factor, we find a singular result when 
$\alpha = D^-, j = 1$ and $\alpha = a,j = 0$. 
We introduce the IR regulator $\delta_{\rm IR}$, which will cancel out against the 
DR term, like in the scalar and fermion cases.
Omitting terms that go to zero with $\delta_{\rm IR}$, we find
\begin{equation}
\begin{split}
  {T_{1,1}^{a,0}}\big|_{\rm IR}&=-\frac14{\hat m_a^2}\int_\rho \rho^3
  \delta m_{a}^2 \left(\frac14+\gamma_E + \ln \frac{\delta_{\rm IR}\rho}{2}\right) \, , 
  \\ 
  {T_{1,1}^{D^-,1}}\big|_{\rm IR}&=-{\hat m_A^2}\int_\rho \rho^3
  \delta m_{A}^2 \left(\frac14+\gamma_E + \ln \frac{\delta_{\rm IR}\rho}{2}\right) \, .
\end{split}   
\end{equation}
For the $T_{2,0}$ traces without the $S''_{\rm o.d.}$ part, we get
\begin{align}
  T_{2,0}^{\alpha,j} = {\rm tr} \left(\frac{1}{-\partial^2}\delta m_\alpha^2\frac{1}{-\partial^2}
  \delta m_\alpha^2\right)
  =\frac{1}{2(\ell_{\alpha,j}+1)^2}\int_\rho \rho^3 \delta m_\alpha^2 \int_0^1 {\rm d}x 
  \ x^{2\ell_{\alpha,j}+3}\delta m_\alpha^2(x\rho) \, .
\end{align}
Finally, we have to calculate the effects of the mixing term $S''_{\rm o.d.}$. 
The only non-vanishing trace where it enters yields
\begin{align}\label{eq:Gaugemix}
  T_{\rm o.d.}^j={\rm tr} \left(\frac{1}{-\partial^2} S''_{\rm o.d.}
  \frac{1}{-\partial^2} S''_{\rm o.d.} \right)
  = \frac2{(j+1)^2}\int_\rho \rho^3\dot m_A \int_0^1 {\rm d}x \ \dot m_A(x\rho)
  \left(x^{2j+4}+x^{2j+2}\right),
\end{align}
which is taken in the Hilbert subspace 
${\cal H}_a^{jm_Am_B} \oplus{\cal H}_{D^+}^{jm_Am_B} \oplus{\cal H}_{D^-}^{jm_Am_B}$. 
In order to get to the final form, it is useful to recall~\eqref{eq:Mod} and the mirror 
result $\frac1{2\dot m_A} S''_{\rm o.d.} |D^\pm,j\rangle = c^\pm |a,j\rangle$. 

Our next step is to sum over $\sigma$ and $j$; for the first trace we have
\begin{equation} \label{eq:T10vectemp}
\begin{split}
  T_{1,0} &= \sum_j d_j^\phi \left( T_{1,0}^{a,j} + T_{1,0}^{D^+,j} + T_{1,0}^{D^-,j} \right) 
  + \sum_j d_j^T \left(T_{1,0}^{T^+,j} + T_{1,0}^{T^-,j}  \right) 
  \\
  &= \frac{1}{2} \int_\rho \rho \ \delta m_a^2 \sum_{j=0}^\infty \frac{d_j^\phi}{j+1} + 
  \frac{1}{2}  \int_\rho \rho \ \delta m_A^2 \left( \sum_{j=0}^\infty \frac{d_j^\phi}{j+2} +
  \sum_{j=1}^\infty \frac{d_j^\phi}{j} + 2 \sum_{j=1}^\infty \frac{d_j^T}{j+1} \right) \, .
\end{split}
\end{equation}
Note that for the NGB field $a$ and for $D^+$ the sum starts at $j = 0$, while for $D^-$ 
and $T^\pm$ it starts at $j = 1$. 
This is a consequence of the group theoretical structure and our choice of $j$.
However, we can relabel the index in each sum and bring the result to a form similar to those 
of scalars and fermions with the following choice
\begin{align} \label{eq:nuchoice}
  \nu &= j + 1 \, ,     & \text{ for }& \ a, D^\pm \, ,
  \\
  \nu &= j \, ,         & \text{ for } & \ T^\pm \, .
\end{align}
After some manipulations we get
\begin{equation}\label{eq:T10vect}
  T_{1,0} = \frac 1 2 \, \sum_{\nu=1}^\infty\left(\nu \int_\rho \rho \,\delta m_a^2
  +2 \, (2\nu+1)\int_\rho \rho \,\delta m_A^2\right) - \frac12\int_\rho \rho\,\delta m_A^2 \, .
\end{equation}
The last term compensates for the finite difference between the series in $j$ 
in~\eqref{eq:T10vectemp} and in $\nu$. 
Summing over $\sigma$ and $j$ for $T_{1,1}$ we have
\begin{align}
  &T_{1,1} = \left. T_{1,1}^{a,0}\right\vert_{\rm IR} + \left. T_{1,1}^{D^-,1}\right\vert_{\rm IR} +
  \frac{1}{4} \int_\rho \rho^3 \hat m_a^2 \delta m_a^2 \sum_{j=1}^\infty \frac{d_j^\phi}{j(j+1)(j+2)} 
  \\ \nonumber
  & + \frac{1}{4} \int_\rho \rho^3 \hat m_A^2 \delta m_A^2 \left(\sum_{j=0}^\infty \frac{d_j^\phi}{(j+1)(j+2)(j+3)}
        +\sum_{j=2}^\infty \frac{d_j^\phi}{(j-1)j(j+1)}
        +\sum_{j=1}^\infty \frac{2 d_j^T}{j(j+1)(j+2)} 
        \right) \, .
\end{align}
Switching from $j$ to $\nu$ using \eqref{eq:nuchoice}, after some algebra we get
\begin{equation} \label{eq:T11vect}
\begin{split}
  T_{1,1} & = \sum_{\nu = 1}^\infty \frac{1}{8\nu} \int_\rho \rho^3 2 \hat m^2_a \delta m_a^2 
  - \frac{1}{8} \int_\rho \rho^3 2 \hat m_a^2 \delta m_a^2 \left( 
  1 + \gamma_E + \ln \frac{\delta_{\rm IR} \rho}{2} \right) 
  \\
  & + \sum_{\nu = 1}^\infty \frac{1}{8\nu} \int_\rho \rho^3 8 \hat m^2_A \delta m_A^2 
  - \frac{1}{8} \int_\rho \rho^3 8 \hat m_A^2 \delta m_A^2 \left( 
  1 + \gamma_E + \ln \frac{\delta_{\rm IR} \rho}{2}  \right) \, .
\end{split}
\end{equation}

The last trace to assemble is
\begin{equation}
\begin{split}
  T_{2,0} & = \frac{1}{2}\int_\rho \rho^3 \delta m_a^2\int_0^1 {\rm d}x \ \delta m_a^2(x\rho) 
  \sum_{j=0}^\infty \frac{d_j^\phi}{(j + 1)^2} x^{2j + 3} 
  \\
  & + \frac{1}{2}\int_\rho \rho^3 \delta m_A^2 \int_0^1 {\rm d}x \ \delta m_A^2(x\rho) 
  \left(\sum_{j=0}^\infty \frac{d_j^\phi}{(j + 2)^2} x^{2j + 5}
  +\sum_{j=1}^\infty \frac{d_j^\phi}{j^2} x^{2j + 1} \right) 
  \\
  & + 2 \int_\rho \rho^3 \dot m_A \int_0^1 {\rm d}x \ \dot m_A(x\rho) 
  \left(\sum_{j=0}^\infty \frac{d_j^\phi}{(j + 1)^2} x^{2j + 4}
  +\sum_{j=1}^\infty \frac{d_j^\phi}{(j+1)^2} x^{2j + 2} \right) 
  \\
  & + \frac{1}{2}\int_\rho \rho^3 \delta m_A^2 \int_0^1 {\rm d}x \ \delta m_A^2(x\rho) 
  \sum_{j=1}^\infty \frac{2 d_j^T}{(j + 1)^2} x^{2j + 3} \, .       
\end{split}
\end{equation}
We switch from $j$ to $\nu$ and perform manipulations analogous to those we used in
the scalar and fermion cases when dealing with $T_{2,0}$. 
We arrive at the form
\begin{align}\label{eq:T20vect}
  T_{2,0} & = \sum_{\nu=1}^\infty \frac{1}{4\nu} \int_\rho \rho^3 
  \left( \delta m_a^2 \right)^2 - \frac{1}{4} \int_\rho \rho^3 \left(\delta m_a^2 \right)^2 
  \\ \nonumber
  &+ \frac{1}{2}\int_\rho \rho^3 \delta m_a^2 \int_0^1{\rm d}x \frac{x^3}{1-x^2}
  \left(\delta m_a^2(x\rho) - \delta m_a^2 \right) 
 + \sum_{\nu=1}^\infty \frac{1}{\nu} \int_\rho \rho^3 \left(\delta m_A^2 \right)^2 - 
  \int_\rho \rho^3 \left(\delta m_A^2 \right)^2 
  \\ \nonumber
  & + 2\int_\rho \rho^3 \delta m_A^2 \int_0^1{\rm d}x 
  \frac{x^3}{1 - x^2} \left( \delta m_A^2(x\rho) - \delta m_A^2 \right) 
  \\ \nonumber
  & + 2 \sum_{\nu=1}^\infty \frac{1}{\nu} \int_\rho \rho^3 
  \dot m_A^2 + \frac{4}{3} \left(-4 + 3 \ln 2 \right) \int_\rho \rho^3 \dot m_A^2 
  + 4 \int_\rho \rho^3 \dot m_A \int_0^1{\rm d}x \frac{x^4}{1-x^2}
  \left(\dot m_A(x\rho) - \dot m_A \right) \, .
\end{align}
This concludes the calculation of the three divergent traces in angular momentum space. 

Next, we move on to momentum space using dimensional regularisation. 
As usual, we have $T_{1,0}^{\rm DR} = 0$. 
The rest naturally splits as $T_{1,1}^{\rm DR} = T_{1,1}^{a,{\rm DR}} + T_{1,1}^{A,{\rm DR}}$ 
and $T_{2,0}^{\rm DR} = T_{2,0}^{a,{\rm DR}} + T_{2,0}^{A,{\rm DR}} + T_{\rm o.d.}^{\rm DR}$.
The contribution from the would-be NGB is given by
$T_{1,1}^{a,{\rm DR}}={\rm RHS}\eqref{eq:T11Reg2}|_{m\to m_a}$ and 
$T_{2,0}^{a,{\rm DR}}={\rm RHS}\eqref{T20regp1}|_{m\to m_a}$.
Altogether, we find
\begin{align} \label{eq:T11DRvect}
  T_{1,1}^{\rm DR} &= \frac{1}{8} \left( \frac{2}{\epsilon} + \ln \frac{\tilde \mu^2}{\delta^2_{\rm IR}} 
  \right) \int_\rho \ \rho^3 \hat m_a^2 \delta m_a^2 + \frac12 \left( \frac{2}{\epsilon}-
  \frac12 + \ln\frac{\tilde\mu^2}{\delta^2_{\rm IR}} \right)\int_{\rho} \rho^3 \hat m_A^2 \delta m_A^2 \, ,
  \\
  \begin{split}
  T_{2,0}^{\rm DR}&= \frac{1}{16\pi^2} \int_k \left\vert\widetilde{\delta m_a^2}(k) \right\vert^2 
  \left( \frac{2}{\epsilon} + 2 + \ln \frac{\tilde \mu^2}{k^2} \right) 
  \\
  &+ \frac{1}{2}\left(\frac{2}{\epsilon}+\frac{3}{2}\right)\int_{\rho}\rho^3
  \big( \delta m_A^2 \big)^2 + \frac{1}{4\pi^2}
  \int_k \left\vert\widetilde{\delta m_A^2}(k)\right \vert^2\ln \frac{\tilde\mu^2}{k^2} 
  \\ 
  &+ 2\left(\frac{1}{\epsilon}+1\right)\int_{\rho} \rho^3\dot m_A^2 +
  \frac{1}{2\pi^2}\int_k\left\vert\tilde m_A(k)\right\vert^2k^2\ln \frac{\tilde\mu^2}{k^2} \, .
\end{split}
\end{align}

As we did in the previous sections, we wrote the above expressions as single $\rho$ integrals
whenever possible, and kept in momentum space the inverse Fourier transforms involving $\ln k^2$. 
Now we can use the identities~\eqref{eq:Fourier} and~\eqref{eq:Fourier2} and obtain 
\begin{align} \label{eq:T20DRvect}
  T_{2,0}^{\rm DR} &= \frac{1}{4} \int_\rho \rho^3 (\delta m_a^2)^2 
  \left( \frac{1}{\epsilon} + \gamma_E + \ln \frac{\tilde \mu \rho}{2} \right) 
  \\ \nonumber
  &+ \frac{1}{2} \int_\rho \rho^3 \delta m_a^2 \int_0^1{\rm d}x \frac{x^3}{1-x^2}
  \left(\delta m_a^2(x\rho) - \delta m_a^2 \right) 
  \\ \nonumber
  &+ \int_\rho \rho^3 (\delta m_A^2)^2 \left( \frac{1}{\epsilon} + \gamma_E 
  -\frac{1}{4} + \ln \frac{\tilde \mu \rho}{2}   \right) 
+ 2\int_\rho \rho^3 \delta m_A^2 \int_0^1{\rm d}x \frac{x^3}{1-x^2}
  \left(\delta m_A^2(x\rho) - \delta m_A^2 \right) 
  \\ \nonumber
  &+ 2 \int_\rho \rho^3 \dot m_A^2 \left( \frac{1}{\epsilon} + \gamma_E -\frac{5}{3} 
  + \ln ( 2\tilde \mu \rho) \right) 
  + 4 \int_\rho \rho^3 \dot m_A \int_0^1{\rm d}x \frac{x^4}{1-x^2}
  \left(\dot m_A(x\rho) - \dot m_A  \right) \, .
\end{align}

We have all the ingredients needed for the regularisation prescription,
\begin{equation} \label{eq:DetSvect}
\begin{split}
 \left. \ln \frac{\det S^{\prime \prime} }{\det {\hat S}^{\prime \prime} }\right
 \vert_{A,a}^{\xi=1} & = \sum_{\nu = 1}^\infty \left( 
 d_\nu^\phi \ln R_\nu^{(Da)} + 2 d_\nu^T \ln R_\nu^{T} \right) 
 \\
 &\qquad \quad - \left( T_{1,0} - T_{1,1} - \frac{1}{2} T_{2,0} \right) + 
  \left( T^{\rm DR}_{1,0} - T^{\rm DR}_{1,1} - \frac{1}{2} T^{\rm DR}_{2,0} \right) \, .
\end{split}
\end{equation}
All the double integrals cancel out, the rest combines into the result given 
in~\eqref{eq:DetS_vect}.

%
%
\section{Outlook} \label{sec:outlook}

In this work we give concise and precise expressions for the regularised determinants 
of different quantum fields, obtained from a well defined double expansion and high 
multipole subtraction.
We believe these simple results are useful for a precise determination of meta-stability
conditions for generic BSM models, especially when additional non-bounce quantum fluctuation 
have a significant impact on the rate.
Here we focused on $D=4$, but the conceptual steps and the power-counting expansions apply
in any $D$ and it would be interesting to derive even more general expressions in generic $D$s,
as in~\cite{Dunne:2005rt} for scalars.
In particular, the $D=3$ case is relevant for computing vacuum decay rates at finite temperature.
%
%
One can also go to higher orders in the subtraction~\cite{Hur:2008yg} and speed up
the multipole convergence.
In our formalism, this would imply computing traces $T_{i,j}$ with a number of insertions 
$i \geq 1$ and $j \geq 2$, that are UV convergent in $D=4$.
Similarly, one can consider using the 
Feynman diagrammatic approach with the double expansion
developed here and generalise it to simplify two loop~\cite{Bezuglov:2018qpq, Ai:2023yce} 
corrections. 
See also~\cite{Carosi:2024lop} for a recent work on two-loop corrections to false vacuum decay rates.

Our final prescription bypasses the Fourier transform and may thus be numerically more 
stable and easier to implement.
This could be useful for expanding on the existing packages~\cite{Ekstedt:2023sqc}, 
perhaps building on the polygonal bounce~\cite{Guada:2018jek} framework and 
its~\texttt{FindBounce}~\cite{Guada:2020xnz} implementation.
Another potential use of these expressions is to derive exact formul{\ae} for quantum 
fluctuations in BSM cases~\cite{Branchina:2013jra,Branchina:2014usa,Khoury:2021zao}, where the bounce and the scalar determinants are known.
These may include the thin-wall~\cite{Ivanov:2022osf} expansion and its thick 
limits~\cite{Matteini:2024xvg}, as well as the double quartic~\cite{Guada:2020ihz}
or similar setups~\cite{Amariti:2020ntv}.

Generally speaking our work simplifies the form of quantum corrections around a scalar 
instanton.
It might be of interest to consider also more general background objects, such as gauge
instantons~\cite{tHooft:1976snw}, sphalerons~\cite{Klinkhamer:1984di, Kuzmin:1985mm} 
or Q-balls~\cite{Coleman:1985ki, Heeck:2022iky}.
These may find practical applications in QCD~\cite{Dunne:2004cp, Dunne:2004sx, 
Dunne:2005te, Dunne:2006ac}.

%
%
{\bf \textsc{ Acknowledgments.}}
We wish to thank Gašper Košir and Marco Matteini for discussions.
MN is supported by the Slovenian Research Agency under the research core funding 
No.~P1-0035 and in part by the research grants N1-0253 and J1-4389.
The work of LU and MN is partially supported by the grant J1-60026.

\appendix

%
%
\section{Fourier transforms with logarithms} \label{app:FTProofs}

The purpose of this appendix is to prove the key equations~\eqref{eq:Fourier} 
and~\eqref{eq:Fourier2} using two different methods, as we believe that both 
approaches have their virtues. 
Both techniques employ a regularisation method, whose aim is to isolate the UV 
divergences that eventually cancel out. 
The method we employ for~\eqref{eq:Fourier} is more straightforward and uses standard 
distribution-theory reasoning, where a finite length scale $a$ acts as a regulator, 
which is sent to zero in the end.
On the other hand, we prove~\eqref{eq:Fourier2} via analytic continuation techniques,
inspired by $\zeta$-function renormalisation.
It is used to make the trace better behaved in 
the UV, since its modified propagators make it go to zero faster for $k\to\infty$. 

{\bf Proof via regulated distributions.}
Let us start with Eq.~\eqref{eq:Fourier} and rewrite it in the following form,
using integration by parts
\begin{align}\label{eqa:start1} \nonumber
  \int_k \big(\tilde f(|k|)\big)^2\ln k^2 &= -\int_x\nabla_x^2f(|x|)\int_yf(|y|)\int_k
  e^{-i(x-y)k}\frac{\ln k^2}{k^2}
  \\
  &=-\int_x\nabla_x^2f(|x|)\int_{y}f(|y|)\frac{1}{2\pi^2|x-y|^2}\bigg({-}
  \gamma-\frac12\ln\frac{|x-y|^2}{4}\bigg)
  \\ \nonumber
  &=\frac{1}{2\pi^2}\int_{x}f(|x|)\int_{y}f(|y|)\nabla_x^2
  \bigg[\frac{1}{|x-y|^2}\bigg(\gamma+\frac12\ln\frac{|x-y|^2}{4}\bigg)\bigg]\, .
\end{align}
The extra factor of $1/k^2$ in the 4D FT of $\ln k^2/k^2$ makes it 
convergent and straightforward to calculate.
To evaluate the last line, we start by looking for a solution $G(x)$, regular 
at infinity and with $x\in {\bf R}^4$, to the partial differential equation
\begin{equation}\label{eqa:pde}
  \nabla_x^2 G(x) = -\frac{1}{(|x|^2+a^2)^2} \, .
\end{equation}
Here, $a > 0$ is a dimensionful regulator that we send to zero at the end.
Since the equation is $O(4)$ symmetric around $x=0$, $G$ can only depend on $\rho=|x|$.
Integrating \eqref{eqa:pde} once from $0$ to $\rho$, we get
\begin{equation}
  2 \pi^2 \rho^3\frac{{\rm d}G(\rho)}{{\rm d}\rho} = \pi^2 
  \left(\frac{\rho^2}{\rho^2+a^2} + \ln\frac{a^2}{a^2 + \rho^2}\right),
\end{equation}
and after a second integration we find
\begin{equation}
  G(\rho)=\frac{1}{4 \rho^2}\ln\frac{a^2 + \rho^2}{a^2} = \frac{1}{4 \rho^2}
  \left(\ln \rho^2 - \ln a^2 + \ln \frac{a^2 + \rho^2}{\rho^2} \right) \, .
\end{equation}
Isolating the $\ln \rho^2/\rho^2$ term and taking the $\nabla^2$, we can establish
\begin{equation}\label{eqa:magic}
  \nabla_x^2\,\frac{\ln|x-y|^2}{|x-y|^2}=-\frac{4}{(|x-y|^2 + a^2)^2} - 
  4 \pi^2 \ln a^2 \delta^4(x-y) - \nabla_x^2\frac{1}{|x-y|^2}
  \ln\frac{|x-y|^2+a^2}{|x-y|^2} \,,
\end{equation}
where we used~\eqref{eqa:pde} and $\nabla_x^2 {|x-y|^{-2}} = 4 \pi^2 \delta^4(x-y)$. 
The last term is $O(a^2)$ as a distribution, so we drop it.
Applying~\eqref{eqa:magic} to the last line of~\eqref{eqa:start1} we get
\begin{align}\label{eqa:prep}
  \int_k \big(\tilde f(|k|)\big)^2\ln k^2 &= -2\left(\gamma_E + \ln\frac a2\,\right)
  \int_{x}f^2(|x|)-\frac{1}{\pi^2}\int_{x}\int_{y}\frac{f(|x|)f(|y|)}{
  (|x-y|^2+a^2)^2} + O(a^2) \, .
\end{align}
Since the LHS does not depend on $a$, the double integral on the RHS has to carry 
a $\ln a $ term that exactly cancels $-2\ln a \int_x f^2(|x|)$.
We can write it explicitly as
\begin{align}\label{eqa:step1}
  \! \int_{x}\int_{y}\frac{f(|x|)f(|y|)}{(|x-y|^2+a^2)^2}=
  16 \pi^3 \int_{\rho} \rho^3 f(\rho) \int_0^1 \text{d} x x^3 f(\rho x) 
  \int_{\theta} \frac{\sin^2\theta }{(1+x^2-2x\cos\theta+a^2/\rho^2)^2} \, .
\end{align}
From here it becomes clear that the divergence comes from the $a \to 0$, $x = 1$, 
$\theta = 0$ region, where the denominator blows up.
Indeed, setting $f(\rho x) \xrightarrow{x \to 1} f(\rho)$, we can manifestly isolate 
the $\ln a$ dependence,
\begin{align}\label{eqa:integral}
  &\int_0^1 \text{d} x \int_{\theta}
  \frac{\sin^2\theta\, x^3}{\left( 1+x^2-2x\cos\theta+a^2/\rho^2 \right)^2}
  = - \frac\pi 4\left(1 + \ln\frac a \rho \right) + O(a) \,.
\end{align}
Consequently, we can split $f(\rho x)=f(\rho)+\big(f(\rho x)-f(\rho)\big)$
in~\eqref{eqa:step1}, such that the first term with $f(\rho)$ brings out the 
$\ln a$ divergence, which cancels against the one in~\eqref{eqa:prep}.
The remaining term with $f(\rho x)-f(\rho)$ is finite when
$x \to 1$, so we can safely send the regulator $a \to 0$ 
already in the integrand and get~\eqref{eq:Fourier}.

%
%
Eq.~\eqref{eq:Fourier2} can be derived in a similar way by considering the key identity 
\begin{align}
 \int_k \big(\tilde f(|k|)\big)^2k^2\,\ln k^2 &
 =-\int_x\nabla_x^2 \big[\hat x_\mu \dot{f}(|x|) \big ]
 \int_y \hat y_\mu \dot{f}(|y|)\int_ke^{-i(x-y)k}\frac{\ln k^2}{k^2}\,,
\end{align}
to be used as a starting point instead of~\eqref{eqa:start1}. 
One needs also
\begin{equation}
  \int_0^1 \text{d}x \int_{\theta}\frac{\sin^2\theta\,\cos\theta\,x^3}{
  \left( 1 + x^2 - 2x \cos \theta + a^2/\rho^2 \right)^2}
  = - \frac\pi 4 \left( \frac83+\ln\frac a {4 \rho} \right)+O(a) \, .
\end{equation}

\vspace{.3cm}

{\bf Proof via analytic continuation.}
We work in the Hilbert space ${\cal H}$ of square-integrable functions in ${\bf R}^4$, 
and consider the Laplacian operator $\partial^2$ and another operator that acts via
multiplication by $\partial_\mu f(|x|)=\hat x_\mu f'(|x|)$, 
where $f:{\bf R}^+\to {\bf R}$ is any regular function. 
From the point of view of $SO(4)$ rotations, the first operator is a scalar, while 
the second is a vector.
We consider the following trace in ${\cal H}$
\begin{equation}\label{eqa:Trace}
\begin{split}
  T(s)&=\text{Tr}\left(\frac{1}{(-\partial^2)^{1+s}}\, \partial_\mu f(|x|)\,
  \frac{1}{-\partial^2}\, \partial_\mu f(|x|)\right) \, .
\end{split}
\end{equation}
When we let $s \rightarrow 0$, we see that this trace corresponds to the 
$\slashed{\partial} m_\psi$-part of~\eqref{eq:T20fer} for fermions by putting $f=m_\psi$, 
and to~\eqref{eq:Gaugemix} for gauge bosons (setting $f=m_A$). 
However, by playing with the extra parameter $s$ we can keep divergences under 
control: we just need to choose $s$ in an appropriate interval, where the trace is finite. 
Notice that this is not the only way to regulate the trace. 
For example, one could add a power of $s+1$ also to the second propagator, which would
double the pole, but the final result would be the same.

We now follow the same reasoning as in the main text and compute~\eqref{eqa:Trace} in 
momentum and angular momentum space.
For the momentum part, we obtain
\begin{align}
  T(s)&=\int_k\int_p  \frac{(k{-}p)^2\tilde{f}(k{-}p)\tilde{f}(p{-}k)}{(k^2)^{s+1}p^2} 
  =-\frac{\pi}{(4\pi)^{2}}\frac{1}{\Gamma(s{+}1)\Gamma(2{-}s)
  \sin(\pi (2{-}s))}\int_k\frac{\tilde{f}(k)^2}{|k|^{-2+2s}}\nonumber 
  \\
  &=\frac{1}{8}  \left(\frac{1}{s}+1\right)\int_\rho \rho^3 \dot{f}(\rho)^2 -
  \frac{1}{(4\pi)^{2}}\int_k\tilde{f}(k)^2 k^2 \ln k^2+\mathcal{O}(s) \, ,
\label{TAfer}
\end{align}
where we used the fact that there exists a range $s_\text{min}\leq\Re (s)\leq s_\text{max}$
wherein the integrals converge absolutely to get the second equality.
We then expanded for small $s \sim 0$ and used the inverse Fourier transform in the  
terms without the $\ln k^2$. 
Notice that $T(s)$ has a simple $1/s$ pole, signalling the divergence of the original trace.

Let us now compute the same trace in angular momentum space, using the basis 
in~\eqref{angularbasis}
\begin{equation}\label{eqa_trmom}
\begin{split}
  T(s)=&\, \sum_{l,l'} \,\sum_{a, a'} ~\langle l,a|\hat{x}^\mu|l',a'\rangle 
  \langle l',a'|\hat{x}_\mu|l,a\rangle\int_\rho \rho \dot f(\rho)  \int_{\rho'} \rho' \dot f(\rho')
  \\ 
  & \times \int_\lambda\frac{J_{l+1}(\lambda \rho) J_{l+1}(\lambda \rho') }{\lambda^{2s+1}}  
  \int_{\lambda' }\frac{J_{l'+1}(\lambda' \rho) J_{l'+1}(\lambda' \rho') }{\lambda'} \, .
\end{split}
\end{equation}
Let us work out the matrix elements in the first line by using an explicit representation 
of HSHs in terms of $\hat x^a$ coordinates
\begin{align}
  Y_{1,1/2,1/2}(\hat{x})   &= \frac{\hat{x}^1+i \hat{x}^2}{\pi} \, , 
  & 
  Y_{1,-1/2,-1/2}(\hat{x}) &= \frac{\hat{x}^1-i \hat{x}^2}{\pi} \, , \nonumber
  \\
  Y_{1,-1/2,1/2}(\hat{x})  &= \frac{\hat{x}^3+i \hat{x}^4}{\pi} \, , 
  & 
  Y_{1,1/2,-1/2}(\hat{x})  &= \frac{\hat{x}^3-i \hat{x}^4}{\pi} \, ,
\end{align}
and take the sum over polarisations
\begin{align}\label{eqa:matrixY}
  \sum_{a,a'} \langle l,a|\hat{x}^\mu|l',a'\rangle \langle l', a'|\hat{x}_\mu|l,a\rangle
  &=\begin{cases} \frac{\pi^2}{2} \sum_{a,b} \langle l,a| Y_{1,b}^*|l',a+b\rangle 
  \langle l',a+b| Y_{1,b}|l,a\rangle, & l' = l \pm 1 \, ,
  \\
  0,  & \text{else} \, ,
  \end{cases}\nonumber
  \\
  &=\begin{cases}\frac{(l+1)(l+2)}{2} , & l'= l \pm 1 \, ,
  \\
  0,  & \text{else} \, .
  \end{cases}
\end{align}
This result is consistent with the Wigner-Eckart theorem, knowing that $\hat x$ is a 
vector operator.
With \eqref{eqa:matrixY}, changing variable to $\nu=l{+}1$ and integrating over 
$\lambda$ and $\lambda'$, we get
\begin{equation} \label{eqa:HyperG}
  T(s) = \sum_{\nu=1}^\infty  \frac{\Gamma(\nu{-}s)}{2^{1{+}2s}\Gamma(1{+}s)\Gamma(\nu)} \int_\rho \rho^{3+2s} 
  \dot{f}(\rho) \int_0^1 \text{d} x x^{2\nu+2} \dot{f}(\rho x)\ _2F_1(-s,\nu{-}s;1{+}\nu;x^2)\,,
\end{equation}
where the $s$-dependent hypergeometric function ${}_2F_1$ comes from the $\lambda$ integration 
in \eqref{eqa_trmom}. 
To obtain the $1/s$ pole, which is proportional to $\int_\rho \rho^3 \dot{f}(\rho)^2$ as 
in~\eqref{TAfer}, we split~\eqref{eqa:HyperG} into two terms $T(s)=T_{I}(s)+T_{II}(s)$, 
by removing the 0-th order of the Taylor series of $\dot{f}(\rho x)$ around $x\sim1$ ($I$) and
adding it back ($II$), exactly like we did for~\eqref{eqa:step1}. 
The two integrals are treated differently.
For the first term, we expand the integrand around $s\sim 0$ and perform the sum over $\nu$. 
This term is regular for $s\to 0$, and after taking the limit we get
\begin{equation}\label{eqa:T_I}
  \lim_{s\to 0}T_I(s)=\frac{1}{2} \int_\rho \rho^{3} \dot{f}(\rho) 
  \int_0^1 \text{d} x \frac{x^{4}}{1-x^2} \big(\dot{f}(\rho x)-\dot{f}(\rho)\big) \, .
\end{equation}
In order to deal with $T_{II}$, it is convenient to split 
$\int_0^1 \text{d} x x^{2\nu+2} \, _2F_1 =
\int_0^1 \text{d} x x^{2\nu+1}(x-1)\, _2F_1 +
\int_0^1 \text{d} x x^{2\nu+1} \, _2F_1$. 
The reason is that the $1/s$ pole is not contained in the first part, so we can perform 
the expansion around $s\sim 0$ first and then do the sum over $\nu$.
For the second part we have to perform the sum over $\nu$ first and then expand 
around $s\sim 0$, which will give us the $1/s$ pole. 
We obtain
\begin{equation} \label{TBfer}
  T_{II}(s) = \frac{1}{8}\int_\rho \rho^{3} \dot{f}(\rho)^2 \left(\frac{1}{s}-\frac{13}{3}+
  2\gamma+2\ln(2{\rho})\right)+\mathcal{O}(s) \, .
\end{equation}
Equating Eq.~(\ref{TAfer}) and~\eqref{eqa:T_I} +~\eqref{TBfer}, we see that the $1/s$ 
pole cancels out.
We finish by taking the $s \to 0$ limit and recover the desired identity~\eqref{eq:Fourier2}.

Of course one can use analytic continuation to derive~\eqref{eq:Fourier}, too. 
The key is to consider, instead of the vector operator $\partial_\mu f(|x|)$, the scalar 
operator $f(|x|)$.

\section{Analytic formul{\ae} for electroweak vacuum lifetime}\label{app:analytical}

In formulas \eqref{eq:psi_ours} and \eqref{eq:vector_SM} we wrote our results for the top 
and $W, Z$ determinant ratios in terms of regularised series $\Sigma_\psi$ and $\Sigma_{U(1)}$. 
We are going to write here an analytic expression for them in terms of special functions 
${\cal S_\psi}$ and ${\cal S}_\sigma$ from \cite{Chigusa:2018uuj} and a correction term ${\cal D}_A$ 
from~\cite{Baratella:2024hju}. 
For top quarks we found that $\Sigma_\psi(x)={\cal S}_\psi(i\sqrt{x})-3x$, where
\begin{equation}
\begin{split}
  {\cal S}_{\psi}(z) &= \frac{2z}{3}(z^2-1)\big(\ln\Gamma(1+z)-\ln\Gamma(1-z)\big)+\frac{z^2}{3}(z^2-2)
  \gamma_E - \frac{z^2}{9}(z^2+31)
  \\ & + \frac2 3 (1 - 3 z^2) \left( \psi^{(-2)}(1{+}z) + \psi^{(-2)}(1{-}z) \right) + 
  4 z \left(\psi^{(-3)}(1{+}z) - \psi^{(-3)}(1{-}z) \right) 
  \\
  & - 4 \left(\psi^{(-4)}(1{+}z) + \psi^{(-4)}(1{-}z) \right) + 4 \ln A_G + \frac{\zeta(3)}{\pi^2} \, .
\end{split}
\end{equation}
The expression is given in terms of the Glaisher's constant $A_G$ and polygamma functions $\psi^{(n)}(z)$.

To treat $W$ and $Z$ bosons, we need an expression for $\Sigma_{U(1)}(y)$.
We found that it can be written as follows
\begin{align}
  &\Sigma_{U(1)}(y)= \left(2 y - \frac23 y^2 \right) + 2y^2 \bigg(\frac{31}{3}-\pi^2\bigg)+
  \frac12+\frac23\gamma_E+4\ln A_G+\ln\frac{|\lambda|}{8\pi}
  \\
  &+3\, {\cal S}_\sigma\big({-}\tfrac12{+}\tfrac12\sqrt{1-8y}\,\big)+3\ln\big[
  \Gamma\big(\tfrac32-\tfrac12\sqrt{1-8y}\big)\Gamma\big(\tfrac32+\tfrac12\sqrt{1-8y}\big)\big]-\mathcal D_A(y) \, ,\nonumber
\end{align}
where
\begin{align}
  \mathcal S_\sigma(z) &= 
  \frac{z}{6}(1+z)(1+2z)
  \big(
    \ln\Gamma(1+z)-\ln\Gamma(1-z)
  \big)
 +\frac{\gamma_E}{6}z^2(z+1)^2-z-\frac{35}{36}z^2-\frac{z^4}{18} \nonumber \\ &\,
  -
  \left(
    z+z^2+\tfrac{1}{6}
  \right)
  \left(\psi^{(-2)}(1{+}z) + \psi^{(-2)}(1{-}z) \right) 
  +(1+2z)
  \left(\psi^{(-3)}(1{+}z) - \psi^{(-3)}(1{-}z) \right)
  \nonumber \\ &\,
  -2
  \left(\psi^{(-4)}(1{+}z) + \psi^{(-4)}(1{-}z) \right)
  +\frac{1}{2}\ln 2\pi+2\ln A_G+\frac{\zeta(3)}{2\pi^2},
  \label{fnS_B}
\end{align}
\begin{align}
    \mathcal D_A(y)
    &=-2+2y^2(10-\pi^2)+4y(2-\gamma_E)+5\ln2\pi-2\ln (y^{-1}\cos\tfrac{\pi}{2}\sqrt{1-8y})\nonumber\\
    &+ (1-\sqrt{1-8y})\ln\Gamma (\tfrac32-\tfrac12\sqrt{1-8y})+ (1+\sqrt{1-8y})\ln\Gamma (\tfrac32+\tfrac12\sqrt{1-8y})\nonumber\\
    &-2\psi^{(-2)} (\tfrac32-\tfrac12\sqrt{1-8y})-2\psi^{(-2)} (\tfrac32+\tfrac12\sqrt{1-8y}).
\end{align}

\bibliographystyle{apsrev4-2}
\bibliography{references}

\end{document}